\documentclass[%
 reprint,
 amsmath,amssymb,
 aps,
]{revtex4-2}
\usepackage{txfonts}
\usepackage{xspace}
\newcommand{\figwidth}{3in}
\newcommand{\fig}{Fig. }
\newcommand{\eq}{Eq. }
\newcommand{\levy}{L\'evy\xspace}
\newcommand{\um}{\muup\textrm{m}}

\newcommand{\vb}[1]{ {\mathbf #1}}
\newcommand{\rb}{{\mathbf r} }
\newcommand{\lsd}{L_{SD}}
\newcommand{\rhoavg}{\rho_\textrm{avg}}

\usepackage{times}
\usepackage{braket}
\usepackage{color}
\usepackage{graphicx}
\usepackage{dcolumn}
\usepackage{bm}

\begin{document}

\preprint{APS/123-QED}

\title{Active gels, heavy tails, and the cytoskeleton}


\author{Daniel W. Swartz}
\affiliation{Department of Physics \& Astronomy, Johns Hopkins University, Baltimore MD 21218}
\affiliation{Department of Physics, Massachusetts Institute of Technology, Cambridge MA 02139}
\author{Brian A. Camley}%
\affiliation{Department of Physics \& Astronomy, Johns Hopkins University, Baltimore MD 21218}
\affiliation{Department of Biophysics, Johns Hopkins University, Baltimore MD 21218}


\begin{abstract}
The eukaryotic cell's cytoskeleton is a prototypical example of an active material: objects embedded within it are driven by molecular motors acting on the cytoskeleton, leading to anomalous diffusive behavior. Experiments tracking the behavior of cell-attached objects have observed anomalous diffusion with a distribution of displacements that is non-Gaussian, with heavy tails. This has been attributed to ``cytoquakes" or other spatially extended collective effects. We show, using simulations and analytical theory, that a simple continuum active gel model driven by fluctuating force dipoles  naturally creates heavy power-law tails in cytoskeletal displacements. We predict that this power law exponent should depend on the geometry and dimensionality of where force dipoles are distributed through the cell; we find qualitatively different results for force dipoles in a 3D cytoskeleton and a quasi-two-dimensional cortex. We then discuss potential applications of this model both in cells and in synthetic active gels.
\end{abstract}

\maketitle


\section{\label{sec:intro} Introduction}
The cytoskeleton of the eukaryotic cell is a complex material, whose dynamics can to some extent be understood in terms of a gel of semiflexible filaments \cite{broedersz2014modeling,mofrad2009rheology,pegoraro2017mechanical}. However, the dynamics of objects embedded within the cell is not solely controlled by thermal fluctuations, as in a passive gel -- the cell is fundamentally active and out-of-equilibrium, and transport arises from forces exerted by molecular motors, driving deformations of the cytoskeleton \cite{lau2003microrheology,levine2009mechanics,mackintosh2008nonequilibrium,guo2014probing,head2010nonlocal}, complex fluid flows, and surprising changes in transport from equilibrium expectations \cite{stein2021swirling,wolgemuth2020active}. Many of these properties have been recapitulated in minimal ``active gel" experiments, in which a synthetic gel is mixed with molecular motors and ATP \cite{mizuno2007nonequilibrium,toyota2011non,bertrand2012active}. One central observation arising from this work is that even objects embedded in an elastic material can develop a diffusive behavior when active forces are applied to the medium by molecular motors \cite{levine2009mechanics,mackintosh2008nonequilibrium,lau2003microrheology,head2010nonlocal,yasuda2017anomalous,hosaka2017lateral}. Recent works studying dynamics of cell-attached beads and cell cortex attached to an elastic substrate have identified that these active diffusive movements are characterized by ``heavy tails," i.e. a step size distribution (van Hove distribution) $P(\Delta x) \sim \Delta x^{-\alpha}$ that is power-law at large $\Delta x$ \cite{alencar2016non,shi2019dissecting,shi2020dissecting}. Other groups have, in synthetic active gels, observed tails that are heavier than Gaussian, but compatible with exponential tails in $P(\Delta x)$ \cite{toyota2011non}, or even heavier-than-exponential tails \cite{bertrand2012active}. While Gaussian step distributions are a natural expectation due to the Central Limit Theorem, even normal thermal materials may have ``anomalous yet Brownian'' exponential step size distributions \cite{wang2009anomalous} arising from a mechanism of diffusing diffusivity \cite{chechkin2017brownian,chubynsky2014diffusing}. By contrast, heavy power-law tails have been interpreted in the sense of cytoquakes -- that they reflect some rare, coordinated collective movement of the cytoskeleton \cite{alencar2016non,shi2019dissecting}, and simulations provide some support for this idea \cite{floyd2020hessian}. 

In this paper, we show with stochastic simulations and analytical calculations that a minimal active gel model, essentially the one established by Levine and MacKintosh \cite{levine2009mechanics,mackintosh2008nonequilibrium}, naturally creates heavy tails. This arises for a straightforward reason: force dipoles, randomly distributed in the gel, will have a broad range of distances to a tracer point, leading to rare events when force dipoles come close to the tracer, leading to large displacements. Because force dipoles lead to long-range, power law displacements, this range of distances naturally generates a $P(\Delta x)$ with a heavy power law tail, converging to a \levy stable distribution. The emergence of \levy-like distributions from long-range interactions was classically found in the context of electrical dipoles by Holtsmark \cite{holtsmark1919verbreiterung,uchaikin2011chance}, but is also recognized in the statistics of wireless interference \cite{ilow1998analytic,haenggi2009stochastic}, and more recently also found in the statistics of fluids driven by active force dipoles from swimmers \cite{zaid2011levy,zaid2016analytical,kurihara2017non,kanazawa2020loopy} and the mean field theory of plasticity of amorphous solids \cite{lemaitre2007plastic,lin2016mean,parley2020aging}. The key new ideas we show in this paper are that 1) these ideas generalize naturally to viscoelastic materials driven by active force dipoles, 2) we provide explicit formulas for the van Hove distributions, and 3) we show that the expected tail exponent depends strongly on the geometry of the active material -- we expect qualitatively different results in experiments on three-dimensional active gels in vitro and the quasi-two-dimensional cortices in cells. 

\begin{figure}
    \centering
    \includegraphics[width = \figwidth]{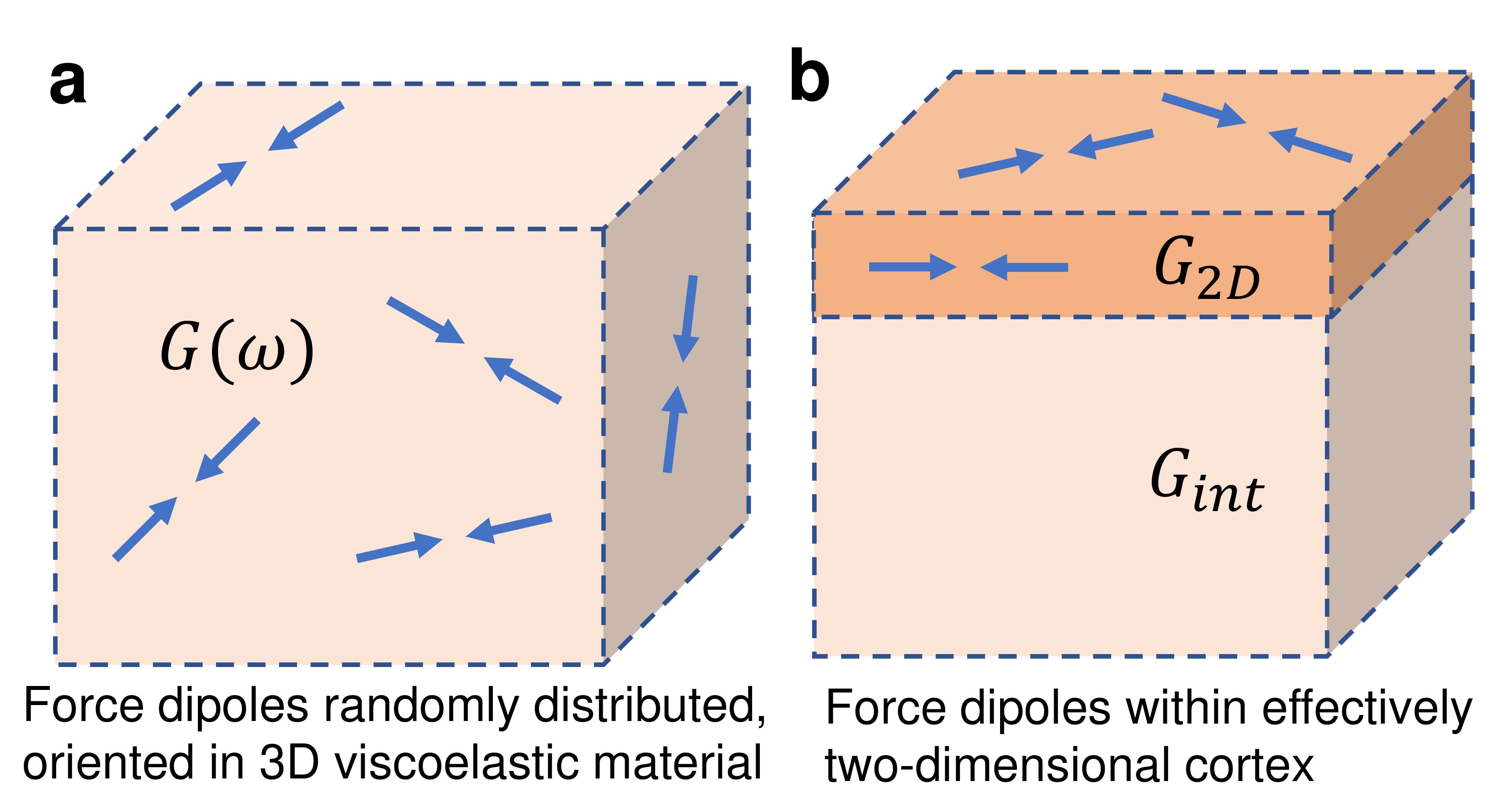}
    \caption{Illustration of geometry. {\bf a:} Three-dimensional geometry. {\bf b:} Quasi-two-dimensional cortex geometry used in Section \ref{sec:quasi2d}.}
    \label{fig:geometry}
\end{figure}
\section{\label{sec:model} Model: Active Gel Driven by Force Dipoles}
We initially describe the cytoskeleton as a three-dimensional isotropic incompressible linear viscoelastic material. The displacement $\vb{u}$ in response to a point force $F(t)$ at the origin is then given by, in Fourier space,
\begin{equation}
    u_i(\rb,\omega) = T_{ij}(\rb,\omega) \vb{F}_j(\omega)
\end{equation}
where the effective Green's function (``Oseen tensor'' in the context of fluid dynamics \cite{doiedwards}) at frequency $\omega$ is, mapping between viscous and viscoelastic systems (see \cite{xu2007correspondence,squires2010fluid}):
\begin{align}
T_{ij}(\rb,\omega) &= \frac{1}{8\pi G(\omega)} \frac{1}{r}\left(\hat{r}_i \hat{r}_j + \delta_{ij}\right)\\
&\equiv \frac{1}{G(\omega)} \widetilde{T}_{ij}(\rb)
\end{align}
We describe our cytoskeleton as being acted on by force dipoles, to represent myosin minifilaments \cite{verkhovsky1995myosin,mackintosh2008nonequilibrium} (\fig \ref{fig:geometry}a). We will begin by treating our force dipoles as two forces with amplitude $F(t)$, separated by a finite distance $d$ along a vector $\hat{b}$; we assume the forces are parallel to $\hat{b}$. The displacement field at the origin is then
\begin{align}
\begin{split}
u_i(0,\omega) = -\sum_n &\left[T_{ij}\left(\vb{r}_n + \frac{\vb{b}_n}{2}\right) \right. \\ 
&\left. -T_{ij}\left(\vb{r}_n-\frac{\vb{b}_n}{2}\right)\right] F_n(\omega)\hat{b}_j
\end{split}
\end{align}
where $\vb{b} = d \hat{b}$ and we have assumed Einstein summation. The displacement can also be written as
\begin{equation} \label{eq:u_ft}
u_i(0,\omega) = \frac{1}{G(\omega)} H_i(\omega)
\end{equation}
where
\begin{equation}\nonumber
H_i(t) = -\sum_n \left[\widetilde{T}_{ij}\left(\vb{r}_n + \frac{\vb{b}_n}{2}\right)-\widetilde{T}_{ij}\left(\vb{r}_n-\frac{\vb{b}_n}{2}\right)\right] F_n(t)\hat{b}_j \label{eq:H_ft}
\end{equation}

\subsection{Time dynamics}
We assume that force dipoles from a population $\rho_\textrm{bulk}$ bind and initiate contraction with a rate $k_{on} \rho_\textrm{bulk}$ per unit volume, i.e. that in a system of volume $L^3$ in a time $\Delta t$, there will be on average $k_{on} \rho_\textrm{bulk} L^3 \Delta t$ force dipoles activated. We simulate this by drawing a random number of force dipoles from a Poisson distribution with mean $k_{on} \rho_\textrm{bulk} L^3 \Delta t$, then generating this number of force dipoles uniformly distributed over the system. We assume that the lifetime of force dipoles is exponentially distributed with mean $\tau$. This corresponds to there being an average density of active force dipoles of $\rhoavg = k_{on} \tau \rho_\textrm{bulk}$. We initialize the system with $\rhoavg L^3$ force dipoles active. Force dipole orientations are chosen uniformly over the sphere. We have also performed simulations with a fixed total number of dipoles, switching on and off, and have noticed no important qualitative differences. 

\subsection{Viscoelasticity}
We compute the values $H(t)$ by the time evolution as above. If the material is viscoelastic, the displacement is then computed by Fast Fourier transforming $H_i(t)$ to $H_i(\omega)$, multiplying by $1/G(\omega)$ and reverse transforming to real space. Because our model for $G(\omega) \sim (i\omega)^\beta$ has vanishing shear viscosity at $\omega \to 0$, we suppress the $\omega = 0$ mode, assuming that the net force over the whole simulation trajectory will average to zero. An alternative regularization would be to assume $G(\omega\to0) = G_0$; we have not noticed important distinctions in the heavy tails in selected tests with this approach. 

\begin{figure}
    \centering
    \includegraphics[width = \figwidth]{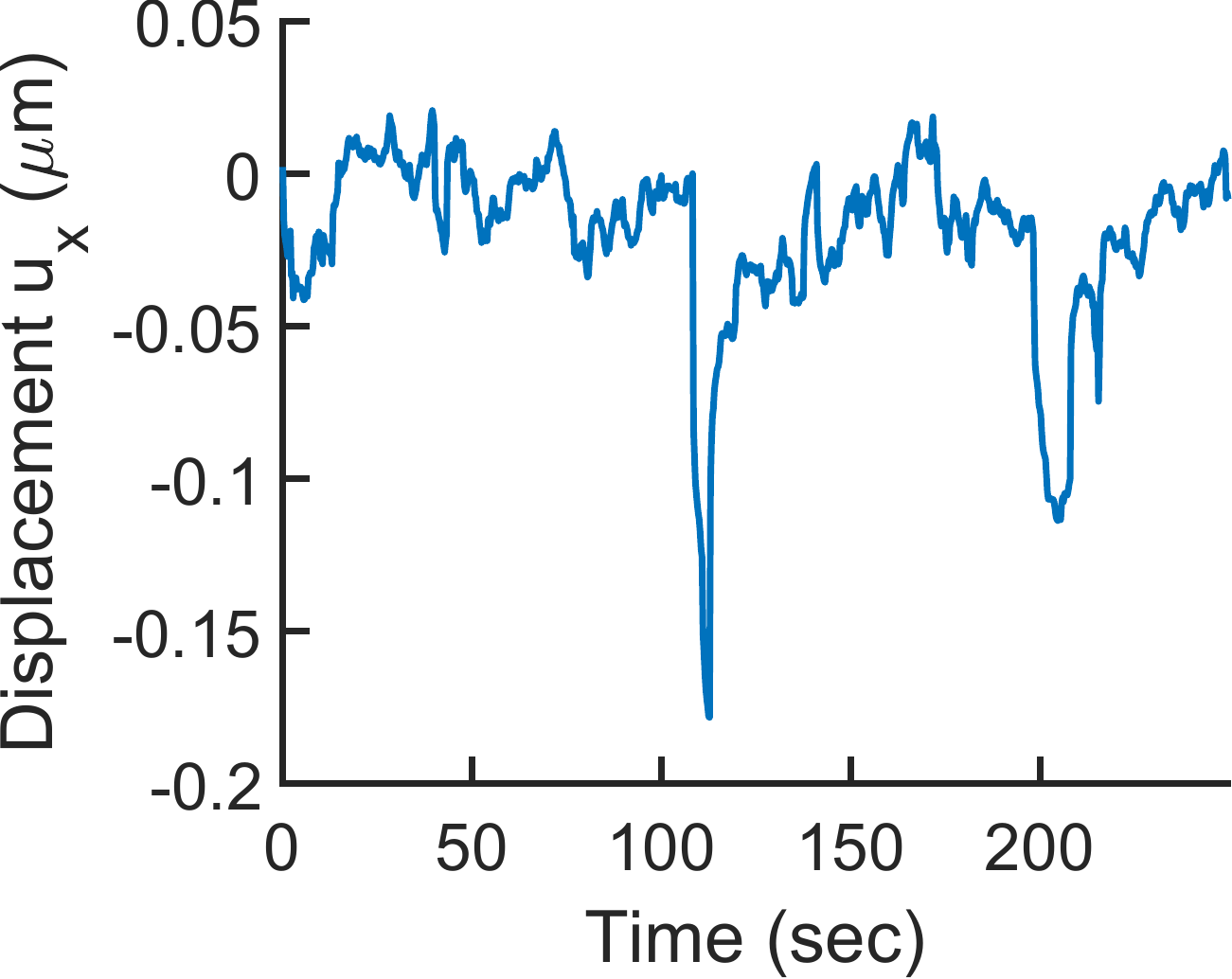}
    \caption{A sample trajectory of the x-axis projection of a dipole-induced random walk, obtained via simulation. Parameters are our default parameter set.}
    \label{fig:trajectory}
\end{figure}

An example trajectory is shown in \fig \ref{fig:trajectory}.

\subsection{Parameters}
We choose the strength of the dipoles as $F = 10$ pN, following \cite{guo2014probing}, and also take our densities to be on the order of $1 \um^{-3}$, as estimated there; $\rhoavg = 0.5 \um^{-3}$ is our default value. We choose our minifilament size to be $d = 0.4$ microns based on \cite{verkhovsky1995myosin}. Unless otherwise stated, we use a power-law rheology of $G(\omega) = G_{scale} (i \omega / \omega_0)^\beta$ with $G_{scale} = 38$ Pa, and $\omega_0 = 10\,\textrm{rad}/s$ \cite{hoffman2006consensus}, and $\beta = 0.17$ as a typical example of a weak power-law consistent with previous results \cite{hoffman2006consensus,guo2014probing,fabry2001scaling,shi2019dissecting}. Our mean motor lifetime $\tau$ is 5 s \cite{guo2014probing}. We use a system size of $L = 80 \um$. Unless otherwise stated, we use a time step of $\Delta t = 0.025 \textrm{s}$, and a total run time of $50,000 \textrm{s}$. 

\section{Simulations show diffusive or superdiffusive dynamics with heavy tails}

\begin{figure}
    \centering
    \includegraphics[width = \figwidth]{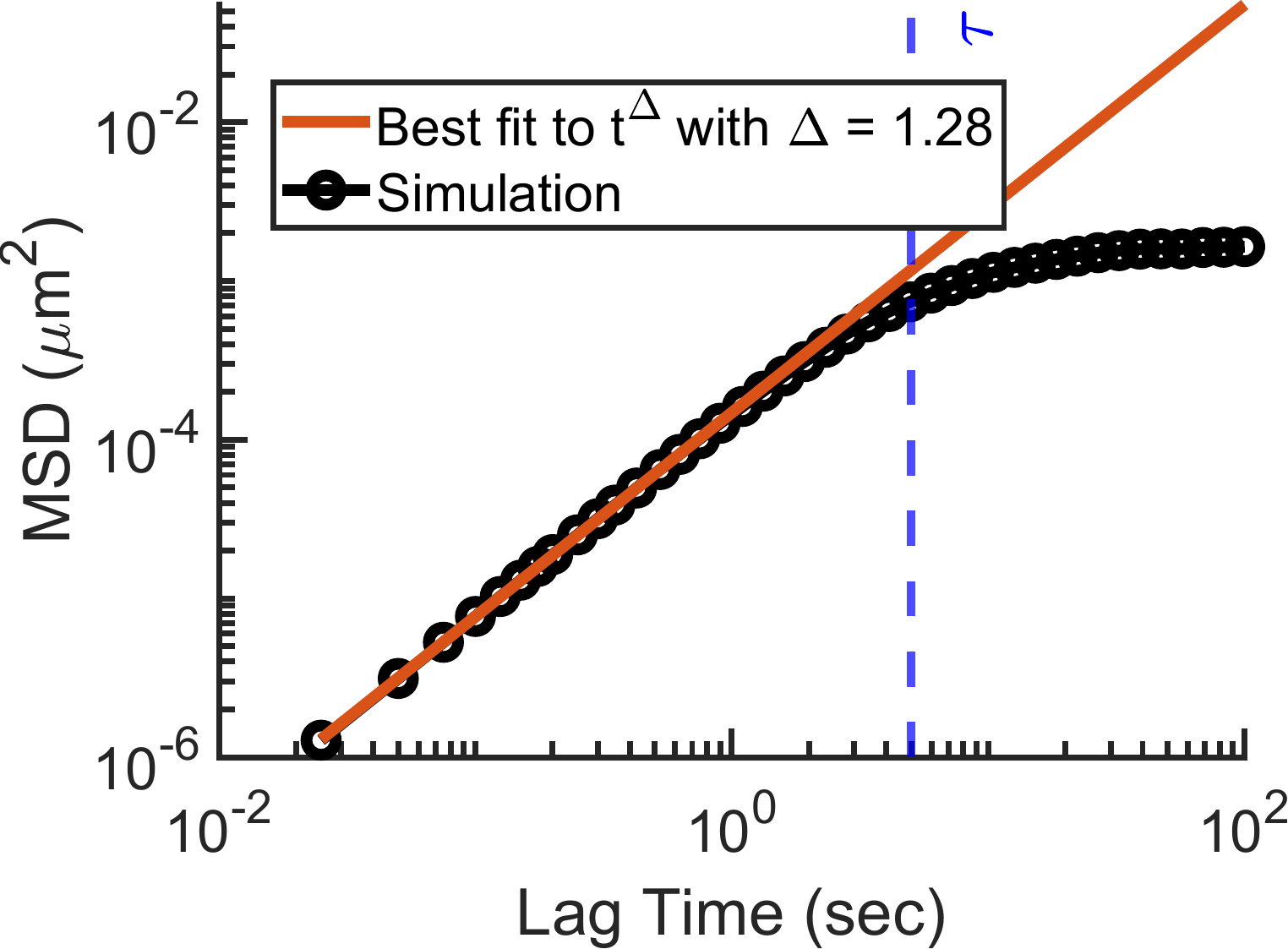}
    \caption{Mean Squared Displacement of displacement trajectory as a function of the lag time. With $G(\omega) \sim (i\omega)^\beta$ ($\beta = 0.17$), we see superdiffusion for times less than $\tau$ (dotted line).}
    \label{fig:msd}
\end{figure}

\begin{figure}
    \centering
    \includegraphics[width = \figwidth]{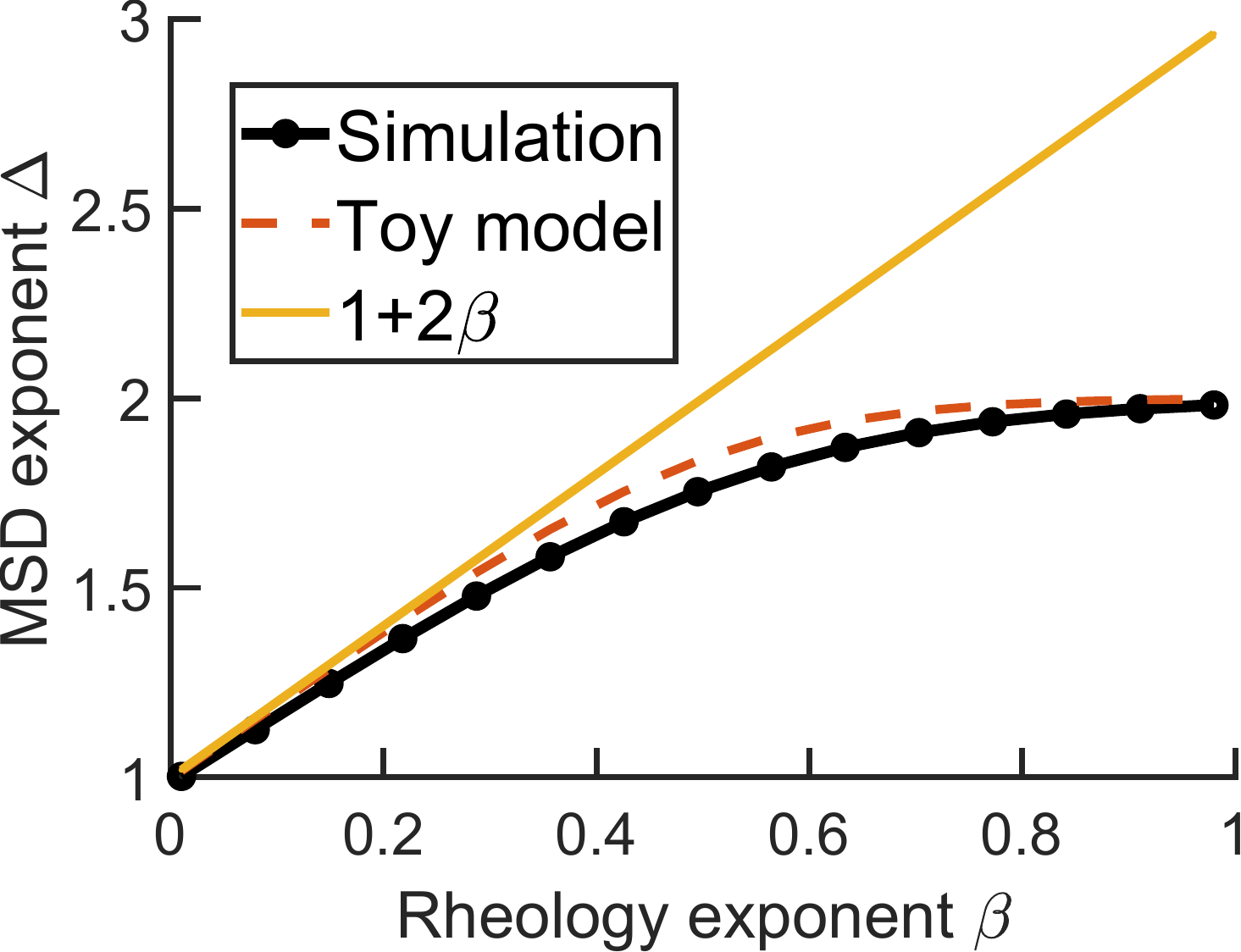}
    \caption{Diffusive exponent $\textrm{MSD}_u \sim t^\Delta$ as a function of rheology exponent $\beta$. Dashed line arises from the simplified model calculation of Appendix \ref{app:toymodel}.} 
    \label{fig:delta_change}
\end{figure}

We find that, as expected from the results of \cite{lau2003microrheology,mackintosh2008nonequilibrium,levine2009mechanics}, that in a purely elastic medium $G(\omega) = G_0$, the corresponding mean-squared displacement of $u_x$, $\textrm{MSD}_u \equiv \langle |u_x(t) - u_x(0)|^2 \rangle$, is diffusive with $\textrm{MSD}_u \sim t^1$ for times shorter than the force dipole correlation time $\tau$, but saturates for $t \gg \tau$. How does this generalize to a viscoelastic material (\fig \ref{fig:msd})? The expectation for a model like this is that in a material with $G(\omega) \sim (i \omega)^\beta$, we should observe $\textrm{MSD}_u \sim t^\Delta$ with $\Delta = 1 + 2 \beta$ \cite{lau2003microrheology,guo2014probing}; this is generally consistent with experimental results for cells measured with small values of $\beta \sim 0.15$ \cite{shi2019dissecting,guo2014probing}. However, we instead observe that $\Delta$ saturates as the power-law exponent is increased (\fig \ref{fig:delta_change}), reaching a maximum of $\Delta = 2$, and showing deviations from $1+2\beta$ even at smaller $\beta$. When the active gel becomes purely viscous at $\beta = 1$, we observe effectively ballistic motion for $t \ll \tau$ -- as would be expected for an object in a viscous fluid driven by a constant pulling force. We can roughly understand the deviations from $t^{1+2\beta}$ by analyzing a simplified version of the active gel model (Appendix \ref{app:toymodel}). These deviations arise for a subtle reason: to derive $t^{1+2\beta}$, Refs. \onlinecite{lau2003microrheology,guo2014probing} assume that the force fluctuations of motors with processivity time $\tau$, which are $\sim \left(\tau^{-2} + \omega^2\right)$, can be treated as $\sim \omega^{-2}$ for times shorter than the processivity time. However, as $\beta$ increases, this approximation fails, as the presence of the processivity becomes a singular perturbation (Appendix \ref{app:toymodel}). 

\begin{figure}
    \centering
    \includegraphics[width = \figwidth]{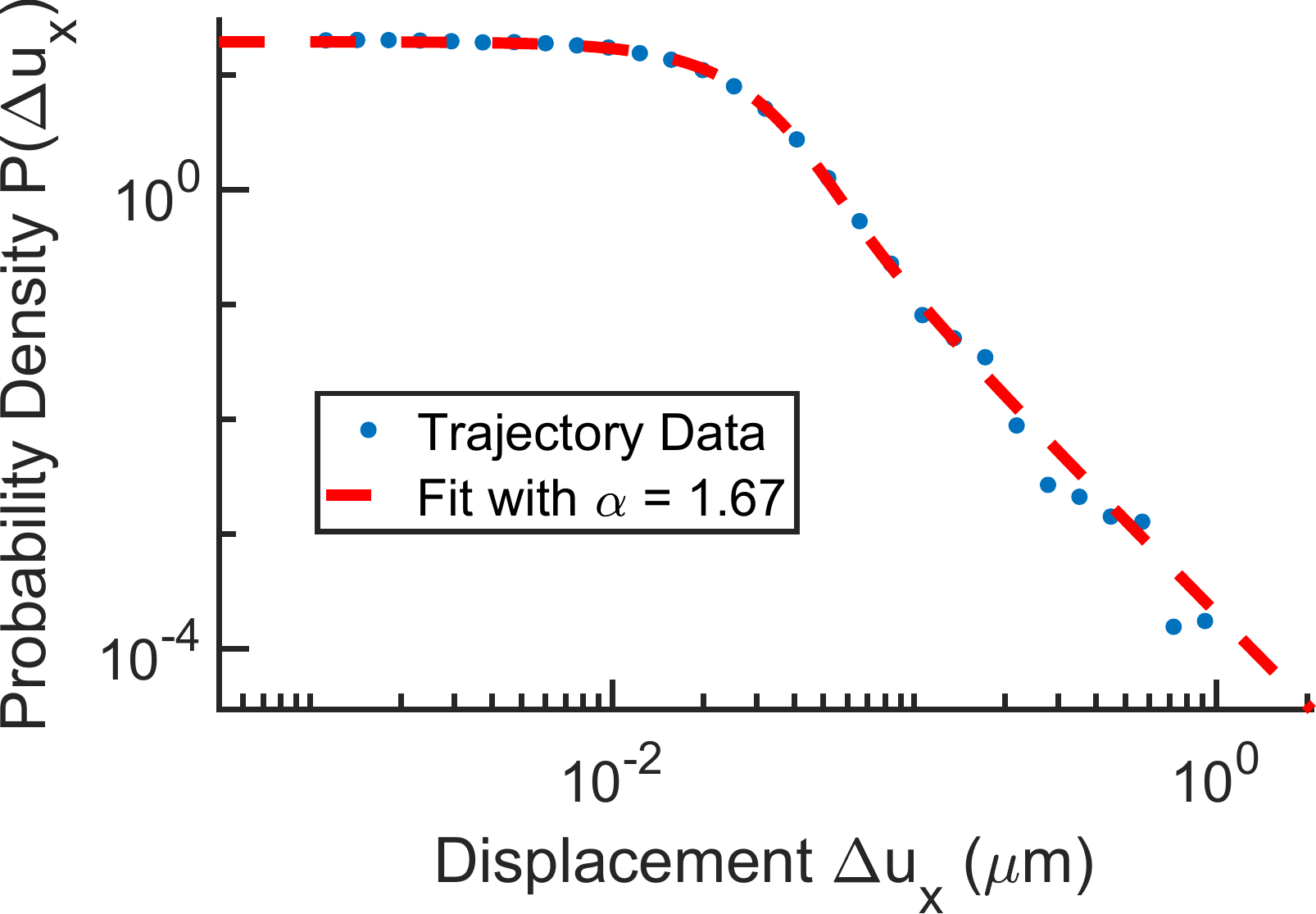}
    \caption{Van Hove correlation plot of the displacement trajectory at time $10$ s, along with a a L\'evy stable distribution, fit via MLE. }
    \label{fig:van_hove}
\end{figure}

From our simulated trajectory $u_x(t)$, we can calculate the distribution of jump sizes at a fixed lag time $P(\Delta u_x)$, with $\Delta u_x = u_x(t+T) - u_x(t)$, the results of which are shown in \ref{fig:van_hove}. This van Hove distribution is fit excellently by a stable distribution with $\alpha \approx 1.67$, corresponding to a power-law tail $P(\Delta u_x) \sim \Delta u_x^{-2.67}$ at large $\Delta u_x$. We note that here, and throughout the paper, we fit to the \levy stable form using the maximum likelihood estimation method (matlab's fitdist). Using the maximum likelihood estimator is the best practice for fitting power-law probability distributions, as many alternatives lead to biases \cite{clauset2009power}. 

\section{\label{sec:intuition} Intuition for why force dipoles generate \levy-Stable distributions}

Our simulations, which have essentially the same assumptions as the classical Levine and MacKintosh theory, include no coupling between force dipoles, and no collective effects. They nonetheless generate a van Hove distribution with heavy tails. Because the displacement of the origin can be written as a sum of independent contributions from our force dipoles, we would naturally expect that the Central Limit theorem would apply and the end result would be Gaussian. Why is this not the case? The answer is that the contribution from each force dipole has a heavy tail and the standard central limit theorem, which assumes that the variance of each individual random variable is finite, does not apply \cite{uchaikin2011chance}. Instead, the sum of many independent heavy-tailed random distributions is a Levy stable distribution. The details of the limit distribution depend strongly on the dimension over which force dipoles are distributed, $D$, and the power-law response. Here is a rough scaling argument: suppose we have force dipoles distributed uniformly in a spherical volume with radius $R$. The probability of an event being a distance $r$ from the origin is given by $P(r) = \int d^Dr\ \delta(|\mathbf{r}|-r) P(\mathbf{r}) = \frac{S_D r^{D-1}}{V_D R^D}$ where $S_D$  and $V_D$ are the surface area and volume respectively of the sphere in $D$ dimensions. To find the distribution of the response $P(u)$, we change variables, using $P(u)du = P(r)dr$, giving 
$    P(u) = P(r)\Big\lvert\frac{dr}{du}\Big\rvert 
    = \frac{S_D r^{D-1}}{V_D R^D} \times \frac{1}{m}u^{-1-\frac{1}{m}} 
    = \frac{S_D u^{\frac{1-D}{m}}}{V_D R^D} \times \frac{1}{m}u^{-1-\frac{1}{m}} 
    = \frac{S_D}{m V_D R^D}u^{-(1+\frac{D}{m})}.$
We see that the response from a single event is distributed with a power law tail $P(u) \sim u^{-(1+D/m)}$. This means that the sum of a collection of $u_i$ will converge to a L\'evy-stable distribution with stability parameter $\alpha = D/m$ \cite{uchaikin2011chance}. (If $D/m \ge 2$, the variance $\langle u^2 \rangle$ is finite and the sum will converge to a Gaussian.) 

As a point force leads to a displacement field that falls off as $1/r$, force dipoles will lead to displacement fields that fall off with $1/r^2$. For our $D = 3$ simulations, we see $m = 2$ and $\alpha = 3/2$. Our best fit to a \levy stable distribution to the data shown in Fig. \ref{fig:van_hove}, though, is $\alpha \approx 1.67$. There are many possible reasons that our simulation could be in disagreement with this basic estimate. First, we have not accounted for the vector nature of the displacement field -- force dipoles create anisotropic displacements. Second, this basic argument does not address the viscoelasticity of the material. Thirdly, in calculating a van Hove distribution, the displacements $u_x(t)$ and $u_x(0)$  are not necessarily composed of a sum of independent heavy-tailed distributions, because the same force dipole may be bound in both cases, leading to correlations. Finally, we have assumed that the dipoles have a finite size $d$, so the displacement field $u$ will not be perfectly $1/r^2$ at length scales close to $d$. In the next section, we show explicitly that the first three objections {\it do not change} the predicted value of $\alpha = 3/2$.

\section{\label{sec:analytic} Analytical calculation of van Hove functions}

We can explicitly show that the van Hove function for a three-dimensional viscoelastic material driven by ideal force dipoles is \levy stable, and compute its properties as a function of lag time. The full calculation is straightforward, following \cite{uchaikin2011chance}, but somewhat lengthy, and we defer it to Appendix \ref{app:vanhove}. 

We calculate the probability of observing a displacement $\Delta u_x(T) = u_x(t + T) - u_x(t)$ at lag time $T$, and find that for a system with our assumed time dynamics, viscoelasticity, and ideal point dipoles that this probability density $p(\Delta u_x)$ is \levy stable. The \levy stable distribution's probability density function does not have an explicit representation, but its characteristic function is simple. For a symmetric \levy stable distribution, this characteristic function is 
\begin{equation}
f_\Delta(k) = \langle e^{-i k \Delta u_x} \rangle = \exp\left[ -\left(\gamma k\right)^\alpha \right]
\end{equation}
Here, $\alpha$ controls the tail exponent of the \levy distribution $p(\Delta u_x) \sim \Delta u_x^{-(\alpha+1)}$, and $\gamma$ sets the scale.
\begin{figure}
    \centering
    \includegraphics[width=0.3\textwidth]{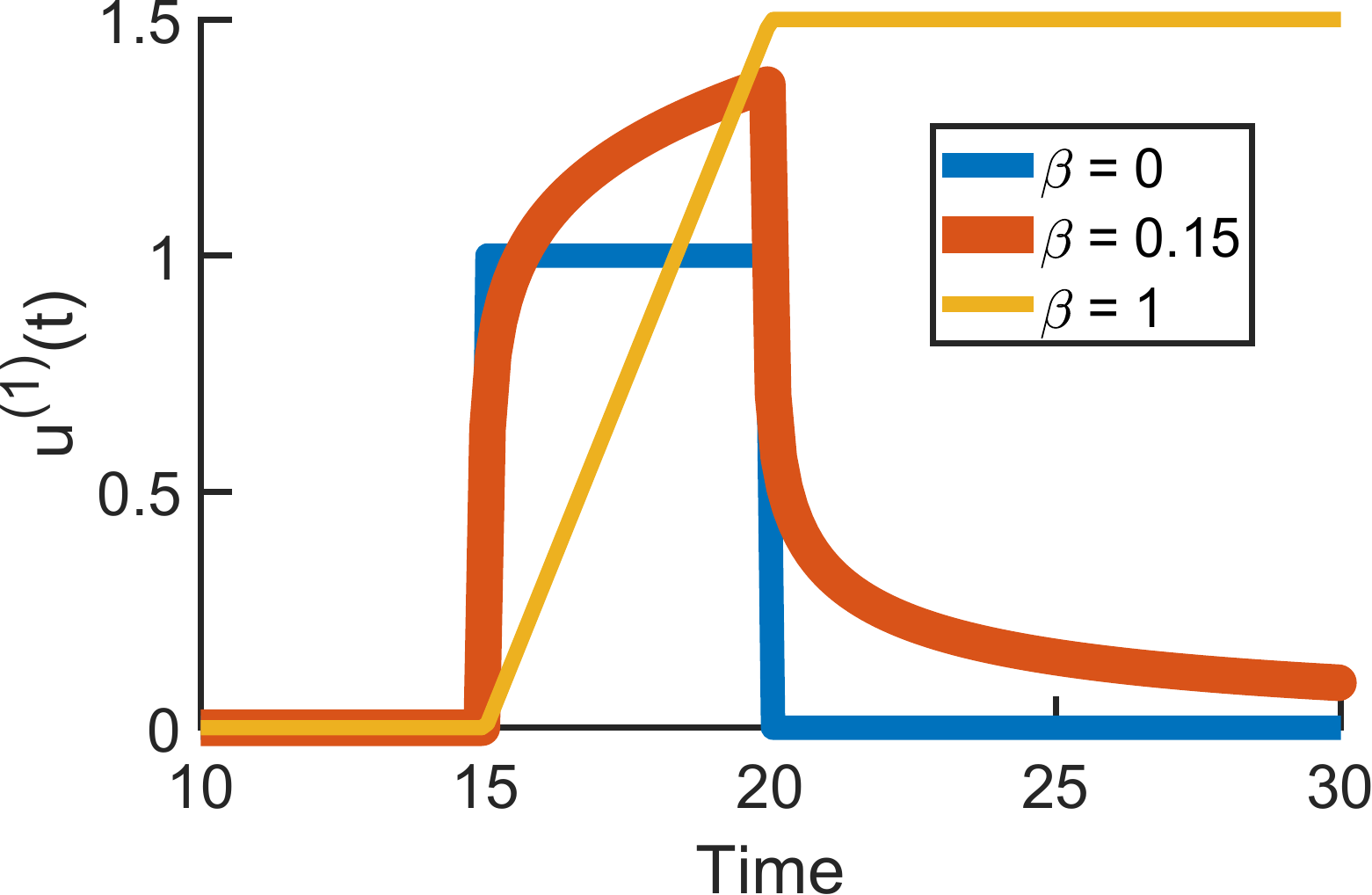}
    \caption{Illustration of response $u^{(1)}$ to a step force with $t_{start} = 15$ and $t_{on} = 5$; the $\beta = 1$ curve has been rescaled so the curves are easy to compare. }
    \label{fig:uone}
\end{figure}

We find that the characteristic function for a viscoelastic material driven by ideal point dipoles is
\begin{widetext}
\begin{align}
\label{eq:levy_visco}
\begin{split}
f_\Delta(k) &= \exp(-k_{on}\rho_\textrm{bulk} \psi_\Delta(k)) \\
    \psi_\Delta(k) &= \frac{B}{60} |k|^{3/2} d^{3/2} \int_0^\infty dt_{on} \tau^{-1} e^{-t_{on}/\tau} \int_{-\infty}^{\infty} dt_{start} |u^{(1)}(t+T;t_{start},t_{on})-u^{(1)}(t;t_{start},t_{on})|^{3/2}
\end{split}
\end{align}
\end{widetext}
where $u^{(1)}(t;t_{start},t_{on})$ is the viscoelastic response at a time $t$ to a a force event of duration $t_{on}$ starting at a time $t_{start}$, i.e. $u^{(1)}(t; t_{start}, t_{on}) = \int dt' \nu(t-t') F(t';t_{start},t_{on})$ where $\nu(t)$ is the inverse Fourier transform of $1/G(\omega)$ (\fig \ref{fig:uone}). $B = \frac{1}{8}\left[2\sqrt{2}+\sqrt{3}\pi+2\sqrt{3}\textrm{arctanh}(\sqrt{2/3}) \right]$.

\eq \ref{eq:levy_visco} shows that the van Hove distribution is \levy-stable with $\alpha = 3/2$ for any lag time, and for any linear viscoelastic material where the integral in \eq \ref{eq:levy_visco} converges. The scale of the distribution, $\gamma$, will change with the lag time $T$, in a complicated way that depends on the rheology. 

We can compute the integral in \eq \ref{eq:levy_visco} explicitly for the simplest case of an elastic material, $G(\omega) = G_0$. Then, we find  $\psi_\Delta(k) = \frac{B}{60} |k|^{3/2} d^{3/2} \left(\frac{F_0}{G_0}\right)^{3/2} \times 2 \tau \left(1-e^{-T/\tau}\right)$. This means that, for an elastic material, we explicitly find:
\begin{align}
\begin{split}
\alpha &= 3/2 \\
\gamma(T) &=\frac{F_0 d}{G_0} \left[ \frac{B}{30} k_{on}\rho_\textrm{bulk} \tau \right]^{2/3} \times \left(1-e^{-T/\tau}\right)^{2/3}
\end{split} \label{eq:gamma_elastic}
\end{align}

\eq \ref{eq:gamma_elastic} is the generalization of computing a mean-squared displacement for a stable distribution: while there is no explicit MSD for a point dipole, because the variance of the stable distribution diverges, $\gamma(T)$ characterizes the scale of the response. At short lag times, we see $\gamma(T) \sim T^{2/3}$, while at times long relative to the motor on time $T \gg \tau$, $\gamma(T)$ saturates; this is akin to the saturation of the MSD observed in our simulations and in earlier work \cite{guo2014probing,mackintosh2008nonequilibrium}.

There are also some simple generalizations of the results \eq \ref{eq:levy_visco} and \eq \ref{eq:gamma_elastic}. For instance, if the dipole size $d$ varies between force dipoles, but this is independent of the dipole's location, the formulas that depend on $d^{3/2}$ will then depend on $\langle |d|^{3/2} \rangle$ (Appendix \ref{app:vanhove}). 

\section{\label{sec:numerical} Numerical simulations: density effects and finite force dipole size}

\begin{figure}
    \centering
    \includegraphics[width = \figwidth]{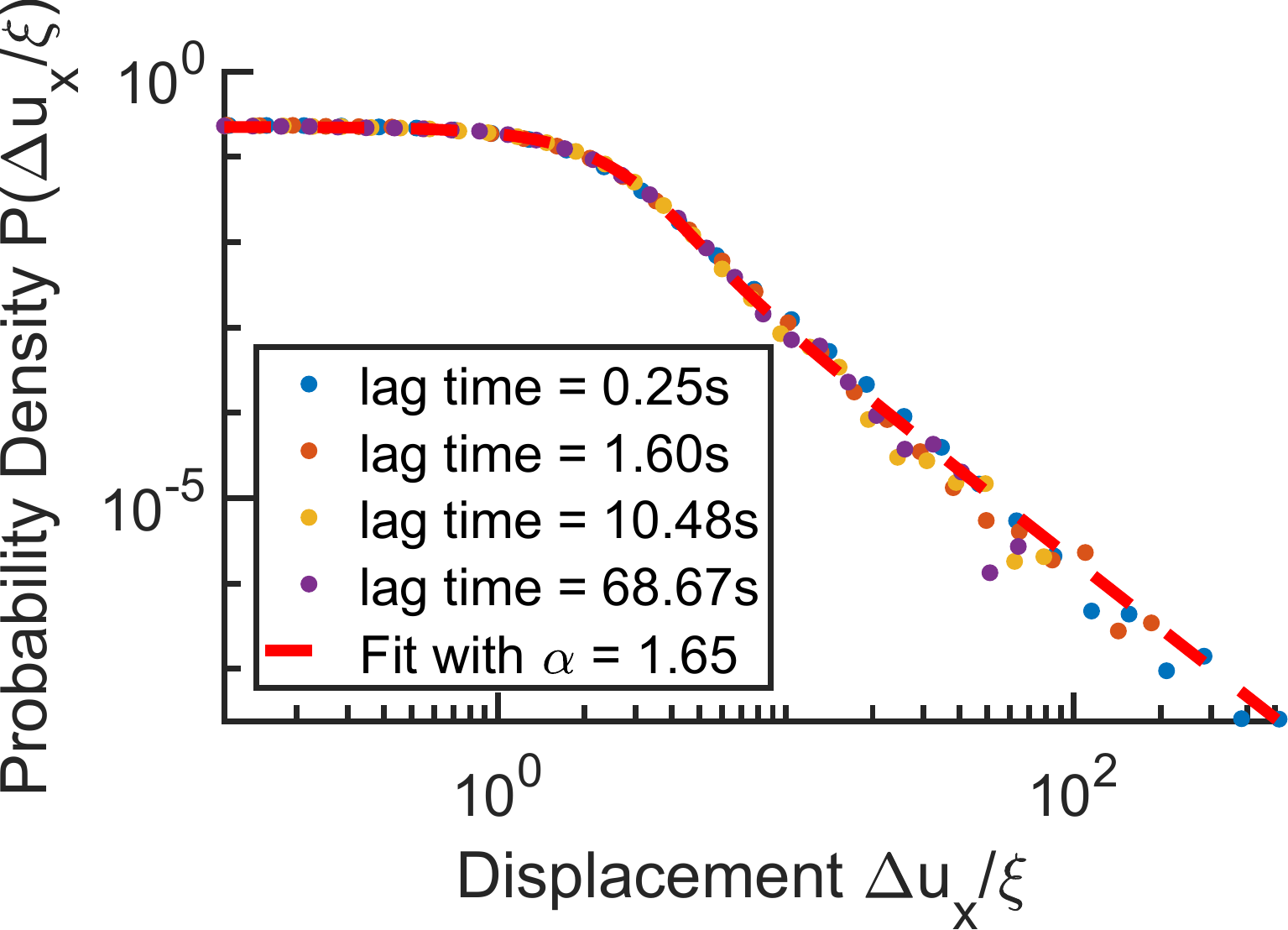}
    \caption{Distribution of displacement $u_x$ at different lag times, rescaled by a lag-time dependent factor $\xi(T) = \exp(\langle\log|\Delta u_x|\rangle)$. }
    \label{fig:rescaledvanhove}
\end{figure}

Our analytical theory predicts that, given an ideal (point) force dipole, the \levy exponent $\alpha = 3/2$, robustly to essentially all factors in the model -- cytoskeletal rheology, density of force dipoles, motor lifetime, etc., contingent on the convergence of the integrals in \eq \ref{eq:levy_visco}. This would predict that the heavy tails should have the exponent $\Delta u^{-2.5}$, relatively universally in many active gels and many cells, and would be robust to many perturbations, as the only crucial elements are the scaling of the response to a dipole and the cell's dimension. However, when we provide reasonable values for the finite force dipole size in our simulations, we see systematic deviations from $\alpha = 3/2$. This is not surprising: when a tracer point is close enough to a force dipole that it sees the individual monopoles, the characteristic displacement changes from $u \sim 1/r^2$ to $u \sim 1/r$ -- suggesting that we should see eventual convergence to a Gaussian, as the variance of our individual forces is finite for $D/m \ge 2$. Nonetheless, we still see heavy tails in our observed van Hove distributions, though we expect them to eventually be cut off at large enough $\Delta u$. (These tails may also be affected if probe or force dipole size leads to force dipoles being excluded from a region near the probe; see Appendix \ref{app:cutoff}.)

We show our simulations for a realistic value of $d = 400 \textrm{nm}$ (a typical minifilament size \cite{verkhovsky1995myosin}) in \fig \ref{fig:rescaledvanhove}. If these distributions were all \levy stable with the same $\alpha$, it would be possible to rescale the van Hove distributions by the scale parameter $\gamma$, and all of $P(\Delta u_x; T)$ would collapse onto a single curve. We choose a slightly different approach, rescaling by $\xi(T) = \exp(\langle\log(|\Delta u_x|)\rangle)$. If $P(\Delta u_x)$ is \levy stable, $\xi$ will be proportional to $\gamma$, but using this approach is more neutral to the choice of distributions. The collapse in \fig \ref{fig:rescaledvanhove} is clear, but not perfect. The van Hove distribution at short times has a heavier tail than that at long lag times. 

\begin{figure}
    \centering
    \includegraphics[width = \linewidth]{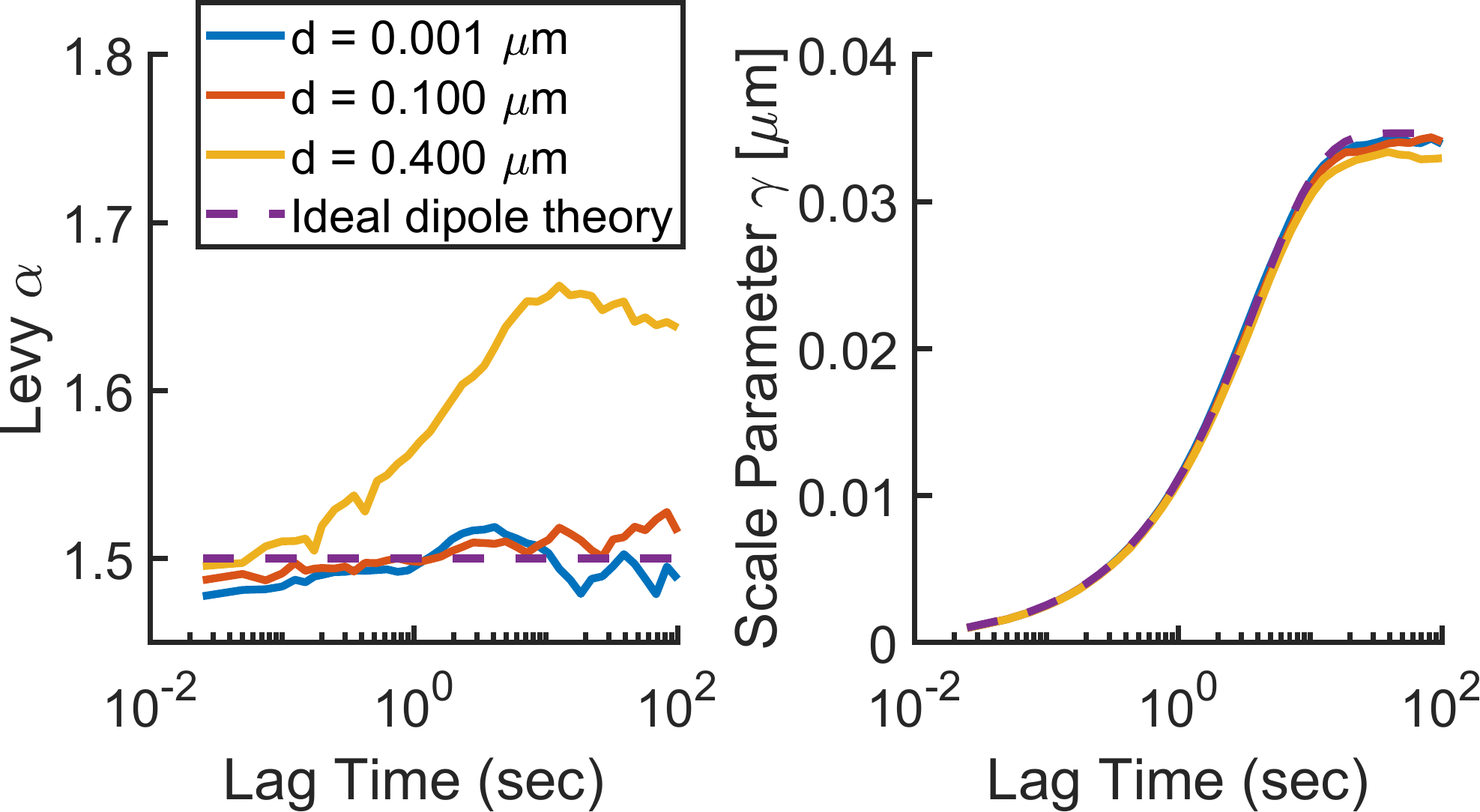}
    \caption{Time dependence of the L\'evy stability parameter $\alpha$ at different dipole sizes. We see at large dipole sizes $\alpha$ grows and becomes dependent on lag time, while at smaller dipole sizes both the tail exponent and scale parameter are given by \eq \ref{eq:gamma_elastic}. In these simulations, $G(\omega) = G_0 = 10$ Pa, and we hold $F d$ constant while varying $d$, so the prediction for $\gamma$ is unchanged. }
    \label{fig:dipolesize}
\end{figure}

Because the van Hove distributions fail to perfectly collapse over all lag times, we expect that the effective tail exponents differ from one lag time to another. We see that this is true in \fig \ref{fig:dipolesize}, finding that $\alpha$ increases systematically over increasing lag time. We also see that our simulations for a perfectly elastic material converge to \eq \ref{eq:gamma_elastic} as the finite dipole size becomes smaller (\fig \ref{fig:dipolesize}).

\begin{figure}
    \centering
    \includegraphics[width = \linewidth]{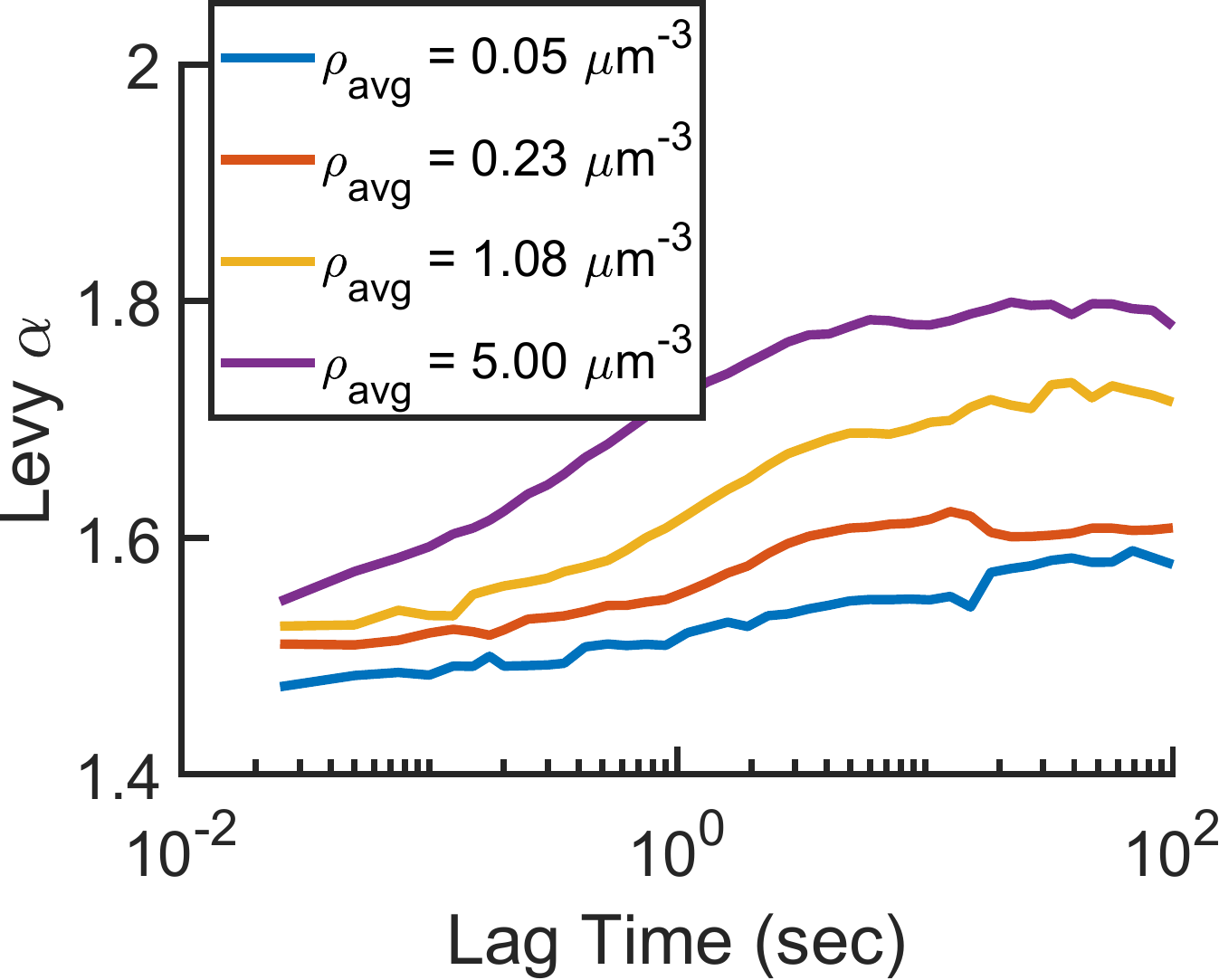}
    \caption{Time dependence of the L\'evy stability parameter $\alpha$ at different densities}
 \label{fig:density}
\end{figure}

Our analytical model for ideal force dipoles (\eq \ref{eq:levy_visco}) predicts that the scale parameter $\gamma$ depends on density $\rhoavg$ -- but the parameter $\alpha$, which controls the heaviness of the tail, should be independent of $\rhoavg$. However, the presence of a finite dipole size alters this dynamic significantly (\fig \ref{fig:density}). We observe that increasing density systematically increases the best-fit $\alpha$ toward its maximum value ($\alpha = 2$, a Gaussian distribution). This would naturally be expected: at larger densities, the typical distance to a nearest force dipole is much smaller, and the displacement induced by this force dipole deviates more strongly from the ideal dipole form. 

\begin{figure}
    \centering
    \includegraphics[width = \linewidth]{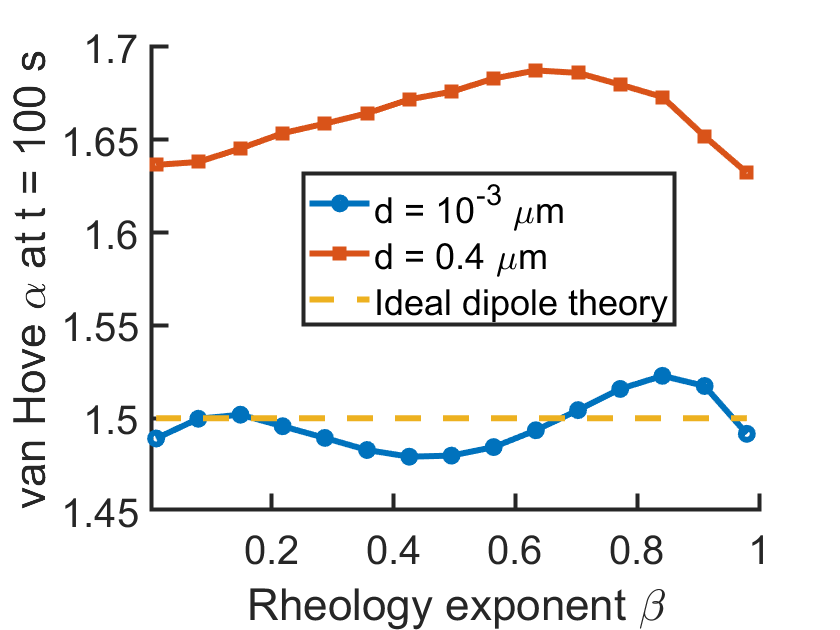}
    \caption{Dependence of the L\'evy stability parameter $\alpha$ on viscoelasticity of the material.}
    \label{fig:betascaling}
\end{figure}

The ideal-dipole calculation (\eq \ref{eq:levy_visco}) suggests that the parameter $\alpha$ should be independent of any details of the viscoelastic response, as long as it is linear. In particular, we predict that it is independent of the exponent $\beta$ of the power-law rheology, $G(\omega) \sim (i \omega)^\beta$. We find, in our simulations, that $\alpha$ is only weakly dependent on $\beta$, even for finite dipole sizes (\fig \ref{fig:betascaling}). 

\section{\label{sec:quasi2d} Effects of quasi-two-dimensional cortex geometry}
Until now, we have treated the case of a three-dimensional linear viscoelastic material - appropriate to understanding the dynamics of tracers embedded within gels in vitro. However, both experiments that show heavy power-law tails are measuring the displacement of tracers at the cell surface \cite{alencar2016non,shi2019dissecting}. The cortex cannot be treated as a three-dimensional material -- typical cortex thicknesses are $\sim 200$ nm \cite{clark2013monitoring,laplaud2020pinching}, while typical minifilament sizes are on the order of 400 nm. We instead move to an opposite extreme, treating the cortex as quasi-two-dimensional (\fig \ref{fig:geometry}b). In this case, we assume that the force dipoles are within the cortex, and planar. If this is the case, we can use the quasi-two-dimensional (``Saffman-Delbr\"uck") Oseen tensor \cite{oppenheimer2009correlated,lubensky1996hydrodynamics,levine2002dynamics,camley2019motion,camley2013diffusion,noruzifar2014calculating}. This Oseen tensor is:
\begin{widetext}
\begin{align}
\begin{split}
T^{SD}_{ij}(\rb) = \frac{1}{4 G_{2D}}\left\{ \left[ H_0 \left(\frac{r}{\lsd}\right) - \frac{H_1\left(\frac{r}{\lsd}\right)}{r/\lsd} - \frac{1}{2}\left(Y_0\left(\frac{r}{\lsd}\right)-Y_2\left(\frac{r}{\lsd}\right)\right) + \frac{2}{\pi (r/\lsd)^2} \right]\delta_{ij} \right. \\ 
\left.- \left[H_0\left(\frac{r}{\lsd}\right) - \frac{2 H_1 \left(\frac{r}{\lsd}\right)}{r/\lsd} + Y_2\left(\frac{r}{\lsd}\right) + \frac{4}{\pi (r/\lsd)^2} \right]\frac{r_i r_j}{r^2} \right\} \label{eq:tsd}
\end{split}
\end{align}
\end{widetext}
where $Y_n$ is the Bessel function of the second kind and $H_n$ the Struve functions. $G_{2D}$ is the 2D shear modulus of the cortex ($G_{2D} \approx G_{\textrm{cortex},3D} h$, where $h$ is the cortex thickness), $\lsd = G_{2D}/G_\textrm{int}$ is the characteristic (Saffman-Delbr\"uck) length, and $G_\textrm{int}$ is the cell interior shear modulus.  

\begin{figure*}
    \centering
    \centerline{\includegraphics[width=\textwidth]{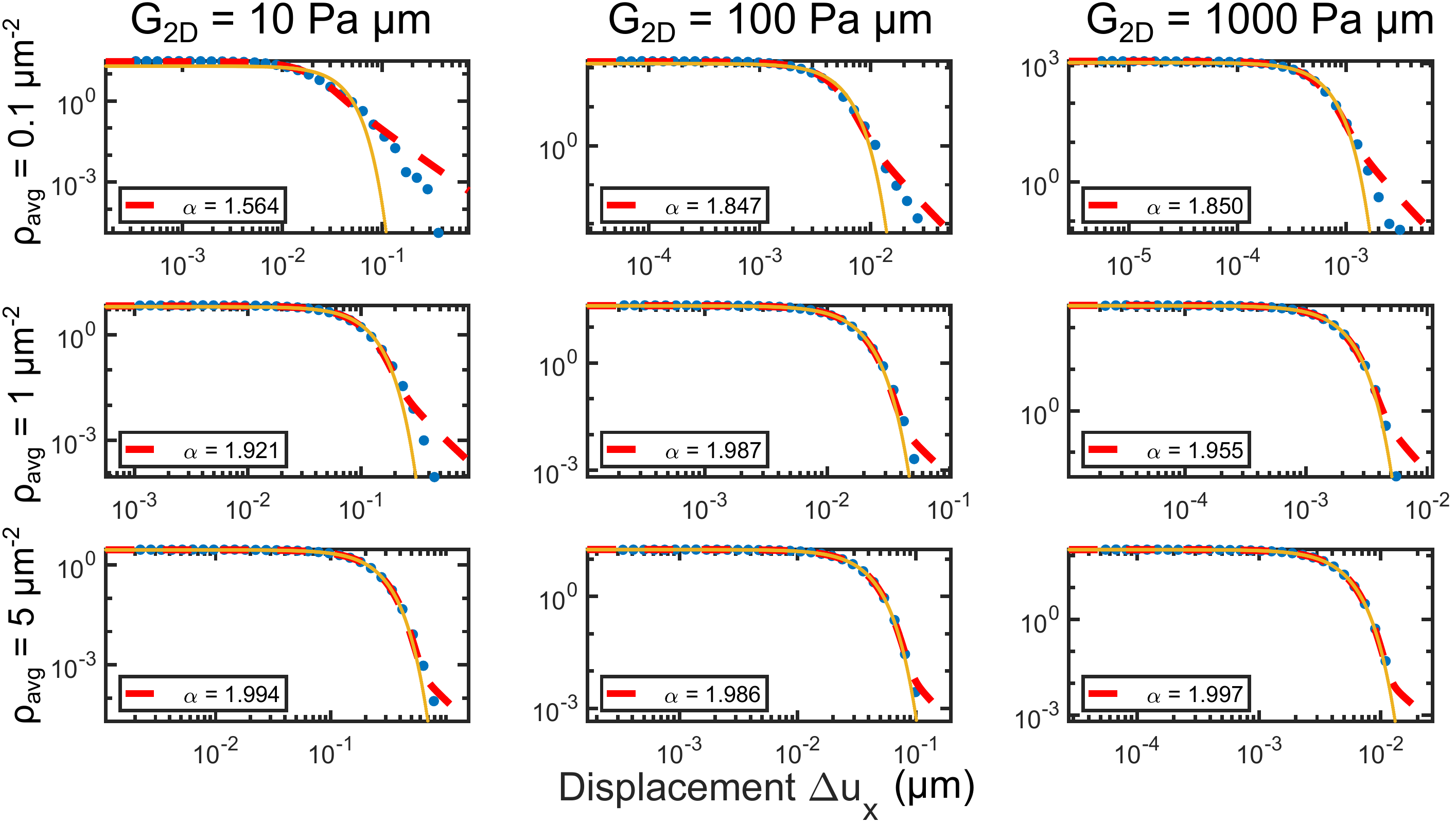}}
    \caption{Van Hove correlation plots $P(\Delta u_x)$ in the quasi-2D geometry at lag time $2 \tau \; (10 \textrm{s})$ for different values of the cortex stiffness (and hence Saffman-Delbr\"uck length) and force dipole density. Blue dots are simulation, red dashed lines are best fit to \levy stable form, yellow solid lines best fit to a Gaussian ($\alpha = 2$).}
    \label{fig:saffvanhove}
\end{figure*}

The form of \eq \ref{eq:tsd} is complex, but can be easily understood in two limits: at distances $r \ll \lsd$, it describes an essentially 2D material, and has a logarithmic dependence on distance,
\begin{equation}
T^{SD}_{ij}(\rb) \approx \frac{1}{4 \pi G_{2D}} \left\{-\left[\ln\left(\frac{r}{2\lsd}\right) + \gamma_E + \frac{1}{2}\right]\delta_{ij} + \frac{r_i r_j}{r^2} \right\}
\end{equation}
while for $r \gg \lsd$, the effect of the cortex can be neglected, and the response is $1/r$, similar to that of a three-dimensional material:
\begin{equation}
T^{SD}_{ij}(\rb) \approx \frac{1}{2 \pi G_{int}} \frac{1}{r} \frac{r_i r_j}{r^2}
\end{equation}

Ideal force dipoles then generate a displacement field that drops off like $u \sim 1/r$ for $r \ll \lsd$ but $u \sim 1/r^2$ for $r \gg \lsd$. Our intuition that we should expect $\alpha = D/m$, then, suggests that if we are primarily seeing dipoles within $\lsd$ of our tracer particle, we should expect $\alpha = 2$ -- i.e. Gaussian van Hoves, but if tracers are primarily separated from force dipoles by distances $r \gg \lsd$, we would expect $\alpha \to 1$ -- i.e. a Cauchy distribution. 

Two-particle microrheology suggests a reasonable range for the cell interior is $\sim 10$ Pa \cite{hoffman2006consensus}, but there is a great deal of range in reported values for cortical stiffness \cite{yamada2000mechanics}-- and both exhibit power-law rheology. 
Unfortunately, our approach for the pure three-dimensional materials will not allow us to study viscoelasticity in the quasi-two-dimensional material. This is because in quasi-2D, the dependence on the shear viscosity can no longer be scaled out as in \eq \ref{eq:u_ft} above -- $T_{ij}^{SD}$ depends on $G(\omega)$ both through an overall scaling and through $\lsd$. For this reason, the way in which a force drives displacement at different times will be significantly more complex \cite{camley2011beyond}. In principle, it would be possible to directly determine $T_{ij}^{SD}(\omega)$ for each force event, and add them up to determine $u(\omega)$ and hence $u(t)$. However, this means that for each event, we must compute $T_{ij}^{SD}$ for all of the frequencies $\omega$ in our problem. Both the number of frequencies needed to capture this response and the number of events scale linearly with simulation time -- leading to a problem that scales quadratically with the total amount of simulation time. We have found that this is intractable for the times required to understand the long tails of our distributions. 

For these reasons, we take the somewhat oversimplified view of representing the cortex and cell interior $G$ values as constant in frequency, and show how the distribution $P(\Delta u_x)$ depends on varying the Saffman-Delbr\"uck length by changing $G_{cortex,3D}$ over plausible values, which range from 10s of Pa to kPa \cite{yamada2000mechanics} (\fig \ref{fig:saffvanhove}). We thus range $G_{2D} = h G_{cortex,3D}$ from $10 \, \textrm{Pa}\, \um$ to $1000 \, \textrm{Pa} \, \um$, and choose $G_\textrm{int} = 10$ Pa as a rough order of magnitude. 

The range of fit $\alpha$ values is consistent with our scaling intuition, with $\alpha$ varying from 1 to 2 (\fig \ref{fig:saffvanhove}, {Appendix \ref{sec:lsd_vary}}). However, over the physically reasonable range of $G_{2D}$ (corresponding to $\lsd = 10-1000 \um$), we find $\alpha$ from $1.56$ to $2$. This is also mediated by the density, as would be expected. The typical distance from our tracer at the origin to a force dipole is $\rhoavg^{-1/2}$. At the largest density we show ($\rhoavg = 5 \um^{-2}$) $\rhoavg^{-1/2} = 0.4 \um \ll \lsd$, and we see that $\alpha$ approaches $2$. (We note that this convergence to $\alpha = 2$ reflects both the finite dipole size and the scaling $T_{ij}^{SD} \sim \ln r$; for smaller dipole sizes, an interesting finite size effect appears; see Appendix \ref{sec:lsd_vary}). However, at the smallest density studied in \fig \ref{fig:saffvanhove}, $\rhoavg^{-1/2}$ is on the order of $3.2 \um$, which is closer to our shortest $\lsd$, allowing $\alpha$ to take on intermediate values. (We also see that $\alpha$ does reach $1$ if we choose the cortex stiffness $G_{2D}$ unphysically small, making $\rhoavg^{-1/2} \gg \lsd$). We note that in \fig \ref{fig:saffvanhove}, our van Hove distributions are not perfectly fit by \levy stable distributions. This is to be expected, given the two-scale dependence on separation as well as the finite dipole size. 

\section{Discussion}
We have developed a simple theory showing that, for a 3D viscoelastic incompressible material driven by ideal force dipoles, that the distribution of step sizes of a tracer embedded in that material are \levy-stable-distributed, with $\alpha = 3/2$, and a straightforward formula for the dependence of the distribution scale $\gamma$ on the lag time. However, our further simulations show several complications to this basic picture -- showing that the finite physical size of force dipoles alter the effective $\alpha$ and create deviations from the \levy-stable picture in the far tails. This finite size can play a significant role in the tails of the van Hove distributions, because these heavy tails inherently probe the short-distance response of the material -- as nicely highlighted by earlier work by Zaid and Mizuno studying the effects of imposing a hard cutoff to power-law responses \cite{zaid2016analytical}. (See also Appendix \ref{app:cutoff}). Our results show how the universal regime of $\alpha = 3/2$ may be reached by looking at smaller densities of force dipoles (\fig \ref{fig:density}). We have assumed that our force dipoles are composed of a pair of force monopoles -- but it is likely that there are also higher-order force multipoles that could arise from the details of minifilament structure. If $\alpha = 3/2$ can be systematically measured in an {\it in vitro} or {\it in vivo} experiment, how corrections arise to this universal value at large densities (and at larger values of $\Delta u_x$, deeper into the tail) may be an interesting way to probe both minifilament force generation and the response of the active gel to force. We have not, for instance, included nonlinear effects of the cytoskeletal network, or how motors may act to stiffen the material \cite{storm2005nonlinear,broedersz2011molecular} -- our theory is solely within the framework of linear viscoelasticity. The interaction of motor-induced stiffening and the high degree of heterogeneity implied by the heavy tails may have interesting interactions: would we expect a heavy-tailed distribution of effective stiffnesses throughout an active material? This might suggest that the difficult problem of sensing stiffness in a disordered fiber network \cite{beroz2017physical,beroz2020physical} would become yet more difficult due to the heavy tails in force-dipole-driven displacements. 

Experimental active gels and cell tracking show a variety of different types of tails. These range from exponential \cite{ toyota2011non} to heavier than exponential \cite{bertrand2012active} and explicitly power-law \cite{alencar2016non,shi2019dissecting}.
Our work is largely inspired by the work of Shi et al. \cite{shi2019dissecting,shi2020dissecting}, who have found evidence that van Hove distributions may be described as \levy stable with $\alpha \approx 1.65-1.76$ \cite{shi2020dissecting}. This is roughly compatible with our results for quasi-two-dimensional cell cortices -- but only if the cortex stiffness is on the lower end of the reported range.
Within our framework, why would these different experiments have different results? We highlight the roles of geometry, motor density, and the force dipole size. For {\it in vitro} gels, the three-dimensional geometry would naively predict $\alpha = 1.5$, but Toyota et al. observe exponential tails \cite{toyota2011non}. For finite-sized dipoles, we predict a transition to Gaussian tails at large densities and large dipole sizes -- when the short-range properties of the force dipoles becomes increasingly relevant. The exponential tails commonly observed may arise from a ``diffusing diffusivity''-type mechanism or local heterogeneity if the mechanism studied here is in the Gaussian regime. By contrast, Alencar et al. observe a power-law tail -- but with an exponent $P(\Delta u_x) \sim \Delta u_x^{-3.62}$ -- though with a different rescaling analysis than the one we use here. This exponent cannot be explained within the framework of a \levy stable distribution, as the variance of this distribution is finite. However, we do see that in our quasi-two-dimensional calculations, we see tails that deviate from a simple \levy stable distribution, and fall off faster than the best-fit \levy stable -- the result of \cite{alencar2016non} could be interpreted in this sense. 

We emphasize that our results do not preclude the existence of cytoquakes. While we may explain the heavy tail in van Hove displacements without invoking any criticality or avalanche-like behavior, we note that there are many independent lines of evidence pointing to a cytoquake-like behavior, including the timing of events and recurrence seen in \cite{alencar2016non} and the spatially-extended correlations observed in \cite{shi2019dissecting}. Understanding the origin of these features and how they can be altered by the heavy tails generated by a simple active gel mechanism studied here is an important open question. 

One striking result of our work is that three-dimensional active gels should have tail exponents that are largely independent of the rheology of the gel (\fig \ref{fig:betascaling}, \eq \ref{eq:levy_visco}), while in the quasi-two-dimensional geometry, it is possible to completely change the tail structure from Gaussian to an extreme value of $\alpha = 1$ simply by changing the shear modulus of the cortex. This result may be testable within the context of reconstituted actin cortices \cite{shah2014symmetry,tan2018self}, though since these cortices have a larger thickness of $\sim 1$ micron, finite thickness effects may also be important \cite{prasad2009two}. In addition, work studying the hydrodynamics of membrane flow have shown that the presence of nearby solid supports, internal friction, and embedded immobile proteins may play a large role in altering the Oseen tensor scaling with distance \cite{oppenheimer2010correlated,oppenheimer2011plane,seki2011diffusion,seki2014diffusion,camley2013diffusion}. Other factors, such as local heterogeneity, could also affect long-range responses, as has been measured in the membrane context \cite{chein2019flow}.  Dynamics in quasi-2D systems may be even more complicated than what we have addressed here: because the Saffman-Delbr\"uck length scale will, in a viscoelastic system, depend on frequency, it is possible that a material may act as effectively two-dimensional at some frequencies -- but three-dimensional at others \cite{camley2011beyond}. We also have not addressed recently observed thickness fluctuations arising from activity within the cortex \cite{laplaud2020pinching}. Altogether, we view the quasi-two-dimensional geometry as one giving a huge amount of control over the scaling of the Oseen tensor $T_{ij}(\rb)$ -- and therefore the predicted tails, making it an ideal testbed for finding a rich zoo of different types of rare events in active gels.

Similarly to the quasi-2D results, within three-dimensional gels, the presence of a boundary will affect the effective elastic response -- significantly altering the tails. This is important in interpreting experiments, like those of Bertrand et al. \cite{bertrand2012active}, where measurements are at the surface of an active gel. Our initial tests of this idea show that the surface effects will create both an anisotropy in tails and alter the overall scaling exponent of the tails, consistent with the observation of anisotropy in tail structure by \cite{bertrand2012active}. Other deviations from our results here may arise from short-range deviations from linear elasticity \cite{ronceray2016fiber,rosakis2015model}, which can change the appropriate power law exponent of $T(\rb)$. Cortex turnover and dynamic thickness variations may also be relevant \cite{laplaud2020pinching}.

\levy statistics are often invoked in questions of foraging and animal travel, in order to explain efficient search \cite{viswanathan2008levy} though this optimality of search is sensitive to many assumptions and the hypothesis remains controversial \cite{pyke2015understanding}, and even the central optimality results have been challenged \cite{levernier2020inverse}. It is possible that heavy tails in cytoskeletal systems may similarly play a role in accelerating rare encounters of cytoskeleton-attached objects with binding targets. However, 
we argue that our results, combined with the earlier work of \cite{zaid2016analytical} in particular, suggest a different view: heavy tails are {\it generic} within our current understanding of active gels. This has important implications for many works studying tracer dynamics, coupling of motors to the matrix, trap escape, etc. in active gels \cite{ben2015modeling,razin2019signatures,woillez2020active,wexler2020dynamics,fodor2015activity}: are these results qualitatively changed by the heavy tails in the noise? 

\section*{Acknowledgments}

We thank Yu Shi, Shankar Sivarajan, John Crocker, and Daniel Reich for extensive discussions about their work \cite{shi2019dissecting,shi2020dissecting}, including the rescaling approach we use in Fig. \ref{fig:rescaledvanhove}, and for sharing preliminary data. BAC acknowledges NSF 1945141. 

\onecolumngrid
\appendix

\section{Deviations from $t^{1+2\beta}$}
\label{app:toymodel}
We can understand the deviations from $t^{1+2\beta}$ by looking at a simplified, zero-dimensional version of the model. If we have a displacement $x(t)$ that is driven by a stochastic motor-driven force $f(t)$, in a material with shear modulus $G(\omega)$, we expect $x(\omega) \sim f(\omega)/G(\omega)$. Then, $\textrm{MSD}_x = \langle |x(t)-x(0)|^2 \rangle$ can be computed as, given $x(t) = \int \frac{d\omega}{2\pi} e^{i\omega t} x(\omega)$
\begin{align}
\langle \left[ x(t)-x(0)\right]^2 \rangle &= \Braket{ \left|\int \frac{d\omega}{2\pi} x(\omega)(e^{i\omega t} - e^{i\omega\times 0}) \right|^2} \\ 
&= \int \frac{d\omega}{2\pi}\int \frac{d\omega'}{2\pi} \langle x(\omega)x(\omega')\rangle \left[ e^{i\omega t} e^{i\omega' t} - e^{i\omega t} - e^{i\omega' t} + 1 \right] \\
&= \int \frac{d\omega}{2\pi}\int \frac{d\omega'}{2\pi} \frac{\langle F(\omega)F(\omega')\rangle}{G(\omega)G(\omega')} \left[ e^{i\omega t} e^{i\omega' t} - e^{i\omega t} - e^{i\omega' t} + 1 \right]
\end{align}
The correlation of a step-function force of amplitude $F_0$, averaged over its possible durations is $\langle F(\omega)F(\omega')\rangle = 2\pi \delta(\omega+\omega') \frac{2 F_0^2 \tau^2}{1 + (\omega \tau)^2}$ \cite{levine2009mechanics}, so we find
\begin{align}
\langle \left[ x(t)-x(0)\right]^2 \rangle &= \int \frac{d\omega}{2\pi} \frac{1}{|G(\omega)|^2} \left[ 1 - e^{i\omega t} - e^{-i\omega t} + 1 \right] \frac{2 F_0^2 \tau^2}{1 + (\omega \tau)^2}\\
&\sim \int d\omega \, |\omega|^{-2\beta}\frac{(1-\cos\omega t)}{1+(\omega \tau)^2}
\end{align}
where we have assumed $G(\omega) \sim \omega^\beta$ in the last line. If we change variables to $w = \omega t$, we find
\begin{align}
\langle \left[ x(t)-x(0)\right]^2 \rangle &\sim t^{-1} t^{-2\beta} \int dw |w|^{-2\beta} \frac{1 - \cos w}{1 + w^2 (\tau/t)^2} \\
&\sim t^{1+2\beta} \int_{-\infty}^{\infty} dw |w|^{-2\beta} \frac{1 - \cos w}{(t/\tau)^2 + w^2}
\end{align}
For $t\ll\tau$ we could neglect the term $t/\tau$ in the denominator of the integrand: this predicts that $\textrm{MSD}_x \sim t^{1+2\beta}$, as noted by \cite{lau2003microrheology}. However, as $\beta$ becomes larger, the resulting integral no longer converges. As $w \to 0$, the integrand with $t = 0$ behaves as $\sim w^{-2\beta}\left[1 - (1-w^2/2)\right]/w^2 \sim w^{-2\beta}$, so the integral will diverge at $t = 0$ if $\beta \ge 1/2$; similarly, the presence of a small but finite $t$ can lead to significant errors even for $\beta < 1/2$. We calculate the integral for finite $t$ numerically using Gauss-Kronrod quadrature (matlab's quadgk), and use this to determine the scaling of $\textrm{MSD}_x$; this is shown in \fig \ref{fig:delta_change}.

\section{Computing distributions of displacements analytically}
\label{app:vanhove}

We have defined above the sum:
\begin{equation}
    H_i(t) =-\sum_n \left[\widetilde{T}_{ij}\left(\vb{r}_n + \frac{\vb{b}_n}{2}\right)-\widetilde{T}_{ij}\left(\vb{r}_n-\frac{\vb{b}_n}{2}\right)\right] F_n(t)\hat{b}_j
\end{equation}
We can extend the results of \cite{uchaikin2011chance} to show explicitly that $H_i$ has a Levy stable distribution, and determine its parameters in the limit of ideal (point) dipoles, i.e. $|b_n| \to 0$. In this limit,
\begin{equation}
    \widetilde{T}_{ij}\left(\vb{r}_n + \frac{\vb{b}_n}{2}\right)-\widetilde{T}_{ij}\left(\vb{r}_n-\frac{\vb{b}_n}{2}\right) \approx b_{n,k} \partial_k \widetilde{T}_{ij}(\vb{r}_n)
\end{equation}
where 
\begin{equation}
    8\pi\partial_k \widetilde{T}_{ij}(\rb) = \frac{1}{r^2} \left[-\hat{r}_k \delta_{ij} -3 \hat{r}_i\hat{r}_j\hat{r}_k + \hat{r}_j \delta_{ik} + \hat{r}_i \delta_{jk}\right] \equiv \frac{1}{r^2} D_{ijk}(\hat{\rb})
\end{equation}
where $\hat{\mathbf{r}} = \mathbf{r}/|\mathbf{r}|$. Then 
\begin{equation}
H_i(t) \approx -\frac{1}{8\pi}\sum_n \frac{1}{r_n^2} b F_n(t)D_{ijk}(\hat{\rb}_n)\hat{b}_j\hat{b}_k 
\end{equation}
Let us then compute the distribution of $H_i(t)$ by computing its characteristic function, $f(k) = \langle e^{-i k H_i} \rangle$. This average is over -- independently -- the positions $\vb{r}_n$, the orientations $\hat{b}$, and the force strength $F_n(t)$. Assuming that these are all independent of one another, and have the same distribution for each $n$, we find
\begin{align}
f(k) &= \left\langle \exp\left(i k \sum_{n=1}^{N} \frac{b F_n}{8\pi} r_n^{-2} D_{ijk}(\hat{r}_n) \hat{b}_{i,n} \hat{b}_{j,n}\right) \right\rangle \\
&= \left\langle \exp\left(i k \frac{b F(t)}{8\pi} r^{-2} D_{ijk}(\hat{r}) \hat{b}_{i} \hat{b}_{j}\right) \right\rangle^N
\end{align}
If we treat the location of the point dipoles as uniformly distributed over a sphere with radius $R$, then let $R \to \infty$, we can give an explicit form for the average over position, writing
\begin{equation}
    f(k) = \left\langle \frac{1}{V_R} \int d^3 r \exp\left(i k \frac{b F(t)}{8\pi} r^{-2} D_{ijk}(\hat{r}) \hat{b}_{i} \hat{b}_{j}\right) \right\rangle^{\rho V_R}
\end{equation}
where we have defined the density of point dipoles so that $\rho V_R = N$. We can then, noting $\frac{1}{V_R} \int d^3r = 1$, write
\begin{equation}
    f(k) = \left[1-\left\langle \frac{1}{V_R} \int d^3 r \left\{1- \exp\left(i k \frac{b F(t)}{8\pi} r^{-2} D_{ijk}(\hat{r}) \hat{b}_{i} \hat{b}_{j}\right)\right\} \right\rangle\right]^{\rho V_R}
\end{equation}
Taking the limit of $V_R \to \infty$, holding $\rho$ constant, noting $\lim_{N\to\infty} (1+x/N)^N = e^{x}$, we see
\begin{equation}
    f(k) = \exp\left(-\rho \psi(k)\right)
\end{equation}
with 
\begin{equation}
    \psi(k) = \left\langle \int d^3 r \left[1- \exp\left(i k \frac{b F(t)}{8\pi} r^{-2} D_{ijk}(\hat{r}) \hat{b}_{i} \hat{b}_{j}\right)\right] \right\rangle
\end{equation}
where the average is over the orientations $\hat{b}$, the magnitudes $|\vb{b}| = d$, and the values of $F(t)$, all assumed to be independent of one another. Note that because $f(k)$ is a Fourier transform of a real function, $f(k) = f^*(-k)$; in addition, because of the symmetry of the problem, $f(k) = f(-k)$ and $f(k)$ is real. This means that $\psi(k)$ must also be real, so we may preemptively take the real part:
\begin{equation}
    \psi(k) = \left\langle \int d^3 r \left[1- \cos\left(\frac{k}{r^2} \frac{b F(t)}{8\pi} D_{ijk}(\hat{r}) \hat{b}_{i} \hat{b}_{j}\right)\right] \right\rangle
\end{equation}
Now we can get the primary scaling of the Levy stable distribution by changing variables in the integral from $\rb$ to $\vb{w} = r\sqrt{|g(k)|}$ where $g(k) = k \frac{b F}{8\pi} D_{ijk}(\hat{r}) \hat{b}_{i} \hat{b}_{j}$. We then get 
\begin{equation}
      \psi(k) = \int d^3 w \langle|g(k)|^{3/2}\rangle \left[1- \cos\left(\pm w^{-2}\right) \right]
\end{equation}
where the $\pm$ arises from the sign of $g(k)$, but will be irrelevant because $\cos(x) = \cos(-x)$. We can break the integral into an angular part and a radial part, which can be evaluated:
\begin{align}
      \psi(k) &= \int d\Omega_w \langle|g(k)|^{3/2}\rangle \int dw w^2 \left[1- \cos\left(w^{-2}\right) \right]\\
      &=\frac{\sqrt{2\pi}}{3}\int d\Omega_w \langle|g(k)|^{3/2}\rangle 
\end{align}
Now, we have to be a little careful about defining the types of averages involved. We have an orientational average over the dipole orientation $\hat{b}$ (i.e. an integral $\frac{1}{4\pi} \int d\Omega_b$) and an independent average over the force strength. So we get 
\begin{equation}
\psi(k) = \frac{\sqrt{2\pi}}{3 (8\pi)^{3/2}}|k|^{3/2} \langle |b F|^{3/2} \rangle \frac{1}{4\pi}\int d\Omega_b\int d\Omega_w |D_{ijk}(\hat{w}) \hat{b}_j \hat{b}_k|^{3/2}
\end{equation}
The orientational factor can be simplified as
\begin{align}
D_{ijk}(\hat{r}) \hat{b}_j \hat{b}_k &=  \left[-\hat{r}_k \delta_{ij} -3 \hat{r}_i\hat{r}_j\hat{r}_k + \hat{r}_j \delta_{ik} + \hat{r}_i \delta_{jk}\right] \hat{b}_j \hat{b}_k \\
&= \left[-\hat{\rb}\cdot\hat{b} \hat{b}_i - 3\hat{r}_i (\hat{r}\cdot\hat{b})^2 + \hat{r}\cdot\hat{b} \hat{b}_i + \hat{r}_i \right]\\
&= r_i \left[1-3(\hat{r}\cdot\hat{b})^2\right]
\end{align}
Because of the isotropy of the problem, we can evaluate the $\Omega_b$ integral with the axis chosen so $\hat{w}$ is along $z$, and $\hat{w}\cdot\hat{b} = \cos \theta_b$:
\begin{align}
     \int d\Omega_b \left|1-3(\hat{w}\cdot\hat{b})^2\right|^{3/2} &= 2\pi \int_{-1}^{1} d\cos\theta_b \left|1-3\cos^2\theta_b\right|^{3/2} \\
     &= 2\pi \frac{1}{8}\left[2\sqrt{2}+\sqrt{3}\pi+2\sqrt{3}\textrm{arctanh}(\sqrt{2/3}) \right] \\
     &\equiv 2\pi B
\end{align}
This leaves
\begin{equation}
\psi(k) = \frac{\sqrt{2\pi}}{3 (8\pi)^{3/2}}|k|^{3/2} B \langle |b F|^{3/2} \rangle \frac{1}{2}\int d\Omega_w |w_i|^{3/2}
\end{equation}
The integral $\int d\Omega_w |w_i|^{3/2}$ must, by symmetry, be independent of $i$, so we can evaluate it with $i$ along $z$ for simplicity, $\int d\Omega_w |w_i|^{3/2} = 2\pi \int d\cos\theta_w |\cos\theta_w|^{3/2} = 2\pi \times 4/5$. Put together and simplified, we get
\begin{equation}
    \psi(k) = \frac{B}{60} \langle |b F|^{3/2} \rangle |k|^{3/2}
\end{equation}
This shows that $H_i$ is a Levy stable distribution with parameters $\alpha = 3/2$ and $\gamma = \left[\frac{B}{60} \langle |b F |^{3/2}\rangle \rho\right]^{2/3}$. Note $(B/60)^{2/3} \approx 0.0866371$. 

\begin{figure}
    \centering
    \includegraphics[width=0.7\linewidth]{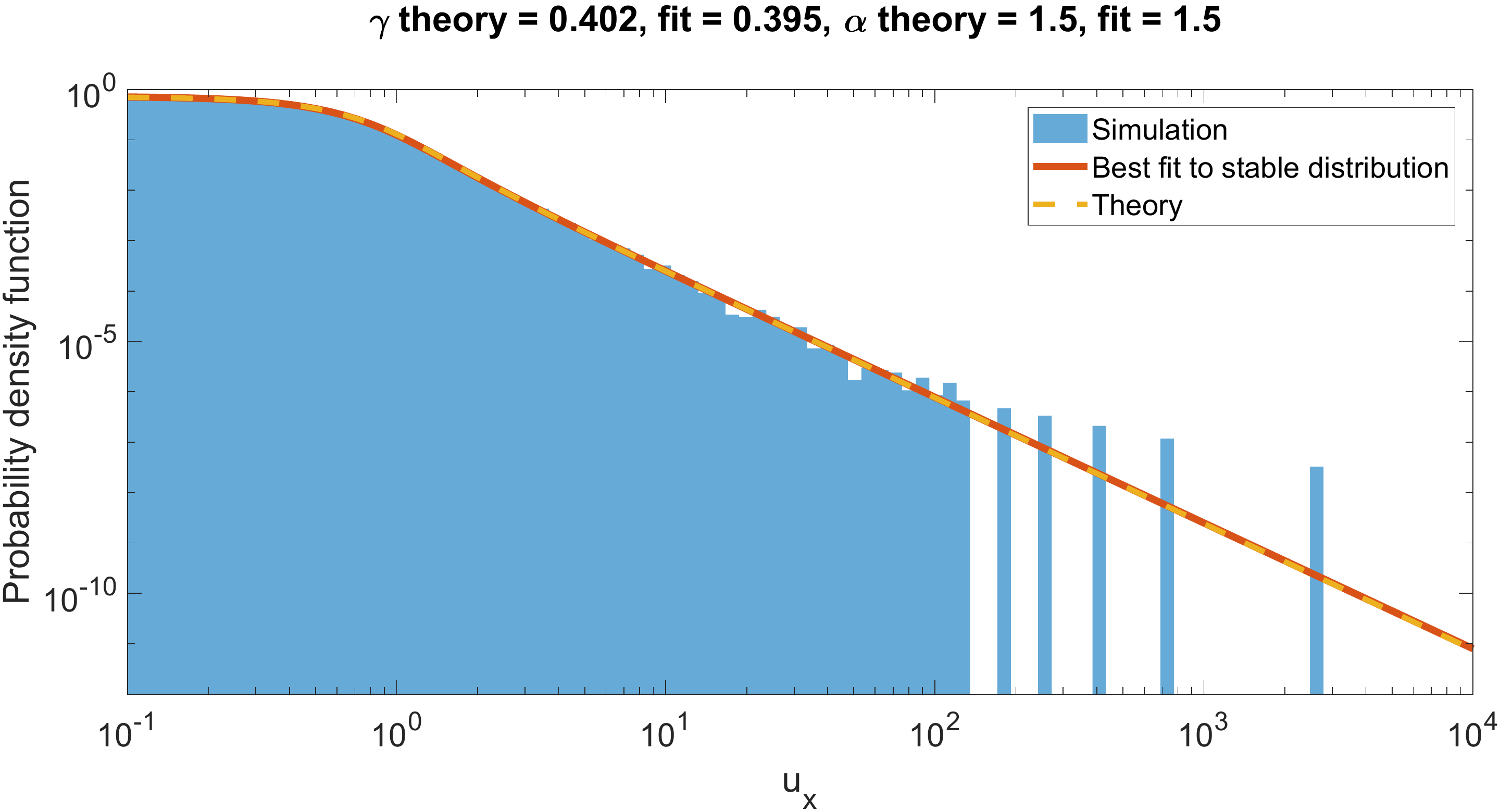}
    \caption{Computing a sum of ideal force dipoles with random distribution gives correct results. This simulation is from $N_{it} = 10^5$ iterations of placing ideal force dipoles with density $\rho = 20$ in a $10\times10\times10$ cube. Here, $Fd$ is chosen to be 1 with probability $p_{on} = 0.5$ and 0 otherwise.}
    \label{fig:ideal}
\end{figure}

We check this in Fig. \ref{fig:ideal}. To study a slightly more general case than our main paper, we choose the force dipole strength $Fd$ to vary randomly, independently of the dipole orientation. 

\subsection{van Hoves and viscoelasticity}

We have to generalize this calculation a little in order to look at the limit of viscoelastic materials. In particular, because of the long-term effects of viscoelastic flow in response to a force, a force in the past leads to displacement at all times. This means that it is essential that we gain a model for how force dipoles are distributed in both time and space. The previous model, with a fixed number of dipoles, makes this a little difficult. However, it's much easier to make this calculation in a model where we have a fixed spatial concentration of dipoles $\rho_\textrm{bulk}$, and describe the rate at which they turn on as $k_{on} \rho_\textrm{bulk}$, with $k_{on}$ as the rate of activation per unit time per unit concentration in the bulk. In this model, the rate of turning on at a time $t$ is constant over the experiment. Generalizing the derivation from above to this case, we get that the characteristic function for $H_i(t)$ is $f(k) = \exp(-\rho_\textrm{bulk} k_{on} \psi)$ with
\begin{equation}
    \psi(k) = \left\langle \int_{-\infty}^{\infty} dt_{start}\int d^3 r \left[1- \exp\left(i k \frac{b F(t;t_{start},t_{on})}{8\pi} r^{-2} D_{ijk}(\hat{r}) \hat{b}_{i} \hat{b}_{j}\right)\right] \right\rangle
\end{equation}
where $t_{start}$ is the time that the force dipole first turns on, $t_{on}$ is its duration. $t$ here is just the time at which $H(t)$ is observed. All of the earlier calculation then carries over, and we find
\begin{equation}
    \psi(k) = \frac{B}{60} |k|^{3/2} \langle |b|^{3/2}\rangle\left\langle \int_{-\infty}^{\infty} dt_{start} |F(t;t_{start},t_{on})|^{3/2}\right\rangle
\end{equation}
where we've assumed that any variation in the dipole length $|b|$ is independent of the strength of the force, dipole position, etc. If the duration of forces has an exponential distribution, as we've assumed in our simulations, we get
\begin{equation}
    \psi(k) = \frac{B}{60} |k|^{3/2} \langle |b|^{3/2}\rangle\int_0^\infty dt_{on} \tau^{-1} e^{-t_{on}/\tau} \int_{-\infty}^{\infty} dt_{start} |F(t;t_{start},t_{on})|^{3/2}
\end{equation}
For a simple force that turns on at time $t_{start}$ and lasts for a time $t_{on}$, we have $F(t) = F_0 \theta(t-t_{start})\theta(t_{start}+t_{on}-t)$, and $\int dt_{start} |F(t;t_{start},t_{on}|^{3/2} = F_0^{3/2} t_{on}$. The integral over $t_{on}$ is then easy to evaluate -- we get the mean on time to be $\tau$, leading to 
\begin{equation}
    \psi(k) = \frac{B}{60} |k|^{3/2} \langle |b|^{3/2}\rangle F_0^{3/2}\tau \label{eq:singletime}
\end{equation}

Crucially, this approach generalizes easily to describing the viscoelastic displacement. Viscoelastic displacement due to a force $F(t;t_{start},t_{on})$ is given by the inverse Fourier transform of Eq. \ref{eq:u_ft}. The linearity of our problem means that 
\begin{equation}
u_i(t) = -\sum_n \left[\widetilde{T}_{ij}\left(\vb{r}_n + \frac{\vb{b}_n}{2}\right)-\widetilde{T}_{ij}\left(\vb{r}_n-\frac{\vb{b}_n}{2}\right)\right] u^{(1)}(t; t_{start,n}, t_{on,n})\hat{b}_j \label{eq:u_realspace}
\end{equation}
where now the sum is over all force dipoles over all time, and 
\begin{equation}
    u^{(1)}(t; t_{start}, t_{on}) = \int dt' \nu(t-t') F(t';t_{start},t_{on})
\end{equation}
and $\nu(t)$ is the Fourier transform of $1/G(\omega)$. $u^{(1)}$ is related to the displacement arising from a single on-off event, and is shown in Fig. \ref{fig:uone} for a power-law material.

By analogy with the earlier results, it is easy to see that the viscoelastic response $u_i$ has a characteristic function $f_u(k) = \exp(-k_{on}\rho \psi_u(k))$ with
\begin{equation}
    \psi_u(k) = \frac{B}{60} |k|^{3/2} \langle |b|^{3/2}\rangle\int_0^\infty dt_{on} \tau^{-1} e^{-t_{on}/\tau} \int_{-\infty}^{\infty} dt_{start} |u^{(1)}(t;t_{start},t_{on})|^{3/2}
\end{equation}
Again, we see that we will get $\alpha = 3/2$ very robustly. However, the prefactor of $k^{3/2}$ above, which controls the scale parameter $\gamma$, can depend on $\beta$. Importantly, however, for our results to be applicable, the integral here must not diverge. This is a serious concern for power-law responses. For times $t \gg t_{start}+t_{on}$, the asymptotic response of $u^{1}(t)$ scales as $t^{\beta-1}$ -- so this integral will only converge if $\frac{3}{2}(\beta-1) < -1$, or $\beta < 1/3$. In practice, this divergence may be regularized if we have $G(\omega) = G_0 + G_{scale} (i\omega/\omega_0)^\beta$ and the long-time response is purely elastic. Within our simulations in the main paper, we are using a finite range of frequencies, so this divergence is also regularized to some extent there. 

Within this language, where our sum in Eq. \ref{eq:u_realspace} is written over all of the force dipole events over all time, it is simple to compute the displacements $u_{i}(t+T) - u_{i}(t)$ as
\begin{equation}
u_i(t+T)-u_i(t) = -\sum_n \left[\widetilde{T}_{ij}\left(\vb{r}_n + \frac{\vb{b}_n}{2}\right)-\widetilde{T}_{ij}\left(\vb{r}_n-\frac{\vb{b}_n}{2}\right)\right] \left[u^{(1)}(t+T; t_{start,n}, t_{on,n})-u^{(1)}(t; t_{start,n}, t_{on,n})\right]\hat{b}_j \label{eq:deltau} \nonumber
\end{equation}
Which then directly shows that the van Hove distributions will have characteristic function 
$f_\Delta(k) = \exp(-k_{on}\rho \psi_\Delta(k))$ with
\begin{equation}
    \psi_\Delta(k) = \frac{B}{60} |k|^{3/2} \langle |b|^{3/2}\rangle\int_0^\infty dt_{on} \tau^{-1} e^{-t_{on}/\tau} \int_{-\infty}^{\infty} dt_{start} |u^{(1)}(t+T;t_{start},t_{on})-u^{(1)}(t;t_{start},t_{on})|^{3/2}
\end{equation}
Again, contingent on the convergence of this integral, we should always see $\alpha = 3/2$ for the ideal dipole response. 

We can evaluate this integral explicitly in the case of a purely elastic system, in which case $u^{(1)}(t;t_{start},t_{on}) = F(t;t_{start},t_{on}) / G_0$. Then 
\begin{align}
    \int_{-\infty}^{\infty} dt_{start}& |u^{(1)}(t+T;t_{start},t_{on})-u^{(1)}(t;t_{start},t_{on})|^{3/2} = \nonumber \\&\left(\frac{F_0}{G_0}\right)^{3/2} \int_{-\infty}^\infty d t_{start} \left|\theta(t+T-t_{start})\theta(t_{start}+t_{on}-t-T) - \theta(t-t_{start})\theta(t_{start}+t_{on}-t)\right|
\end{align}
This integral essentially counts the amount of starting time for which the force dipole is on at only one of $t$ and $t+T$. As $T$ goes to zero, it will vanish -- but it will reach a maximum when $T > t_{on}$. We can evaluate this as 
\begin{equation}
    \int_{-\infty}^\infty d t_{start} \left|\theta(t+T-t_{start})\theta(t_{start}+t_{on}-t-T) - \theta(t-t_{start})\theta(t_{start}+t_{on}-t)\right| = \begin{cases} 
      2 T & T < t_{on} \\
      2 t_{on} & T \ge t_{on}
   \end{cases}
\end{equation}
Computing the average over $t_{on}$ is simple, and we get
\begin{equation}
    \psi_\Delta(k) = \frac{B}{60} |k|^{3/2} \langle |b|^{3/2} \rangle \left(\frac{F_0}{G_0}\right)^{3/2} \times 2 \tau \left(1-e^{-T/\tau}\right)
\end{equation}
Note that the asymptotic limits of this are reasonable -- for $T \ll \tau$, we have that the width of the distribution is vanishing. For $T \gg \tau$, we have a value that is twice that given in Eq. \ref{eq:singletime}. This is because we are effectively summing over twice the density -- because there is no correlation between force dipoles on at time $t$ and at time $t + T$. 

\section{Varying Saffman-Delbr\"uck lengths and system size effects in the quasi-2D geometry}
\label{sec:lsd_vary}

\begin{figure}[h!]
    \centering
    \includegraphics[width=0.6\linewidth]{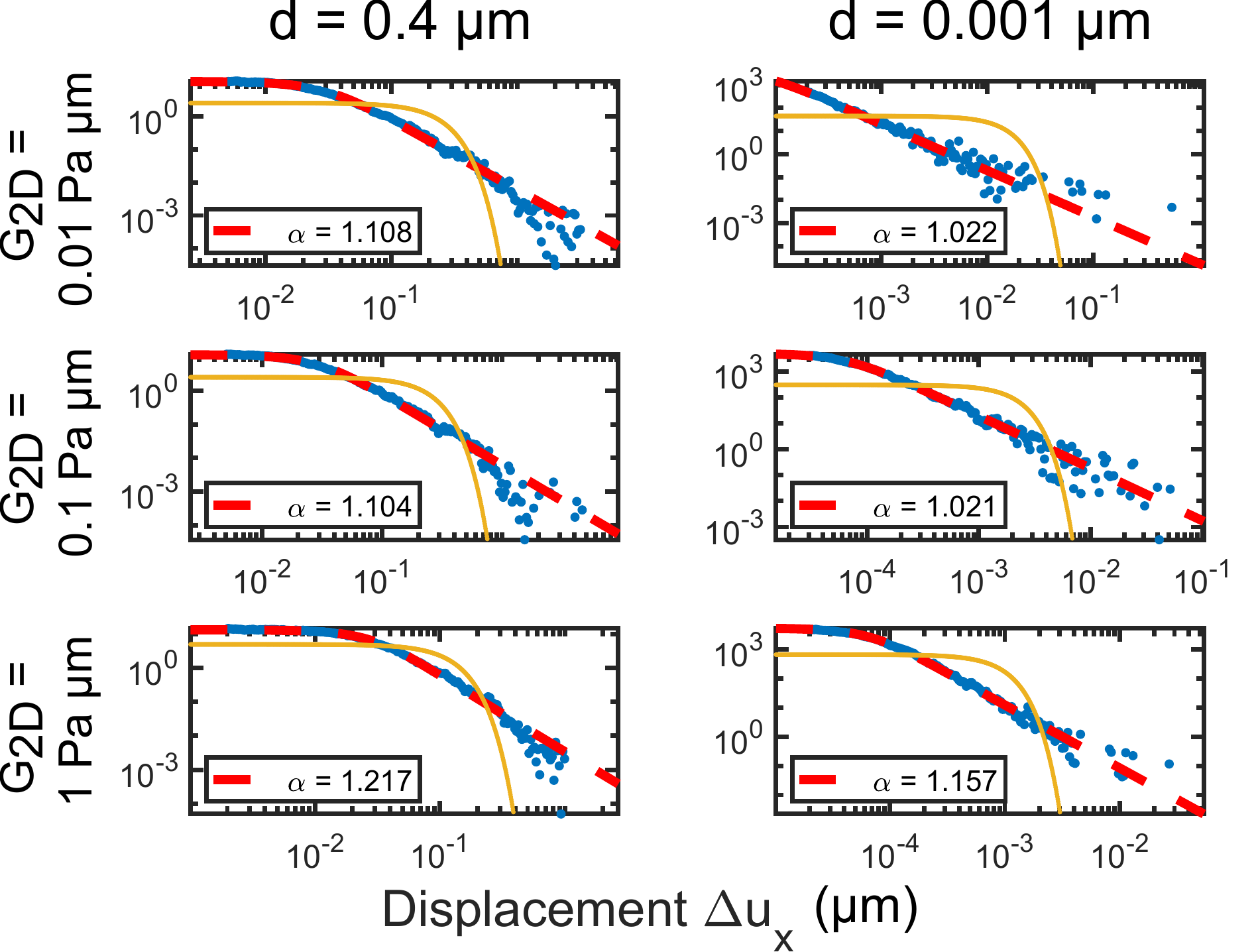}
    \caption{We can capture the limit of $\alpha = 1$ by moving to unphysically small values of $G_{2D}$. As in \fig \ref{fig:saffvanhove} in the main paper, blue dots are simulation data, dashed red line is the best \levy stable fit, and the yellow solid line is the best fit to the Gaussian form. Van Hove plots at lag time $2 \tau \; (10 \textrm{s})$.}
    \label{fig:unphysical}
\end{figure}

In the main text, we showed that it was possible by varying the two-dimension cortex shear modulus $G_{2D}$ over a reasonable physical range of $G_{2D} = 10, 100, 1000 \textrm{Pa} \, \um$ that we found values of $\alpha$ from around $1.5$ to $2$. However, our scaling analysis suggests that we should be able to reach values of $\alpha = 1$ in the limit where $\lsd$ is small relative to typical spacings, and the Oseen tensor for a force monopole becomes $1/r$. We show that this is the case in \fig \ref{fig:unphysical}. We could also reach this limit by increasing the interior shear modulus $G_\textrm{int}$. As we see in the three-dimensional case, with a realistic-sized dipole, there are some small deviations from $\alpha = 1$ which disappear in the limit of $d \to 0$. 

\begin{figure}[h!]
    \centering
    \includegraphics[width=0.6\linewidth]{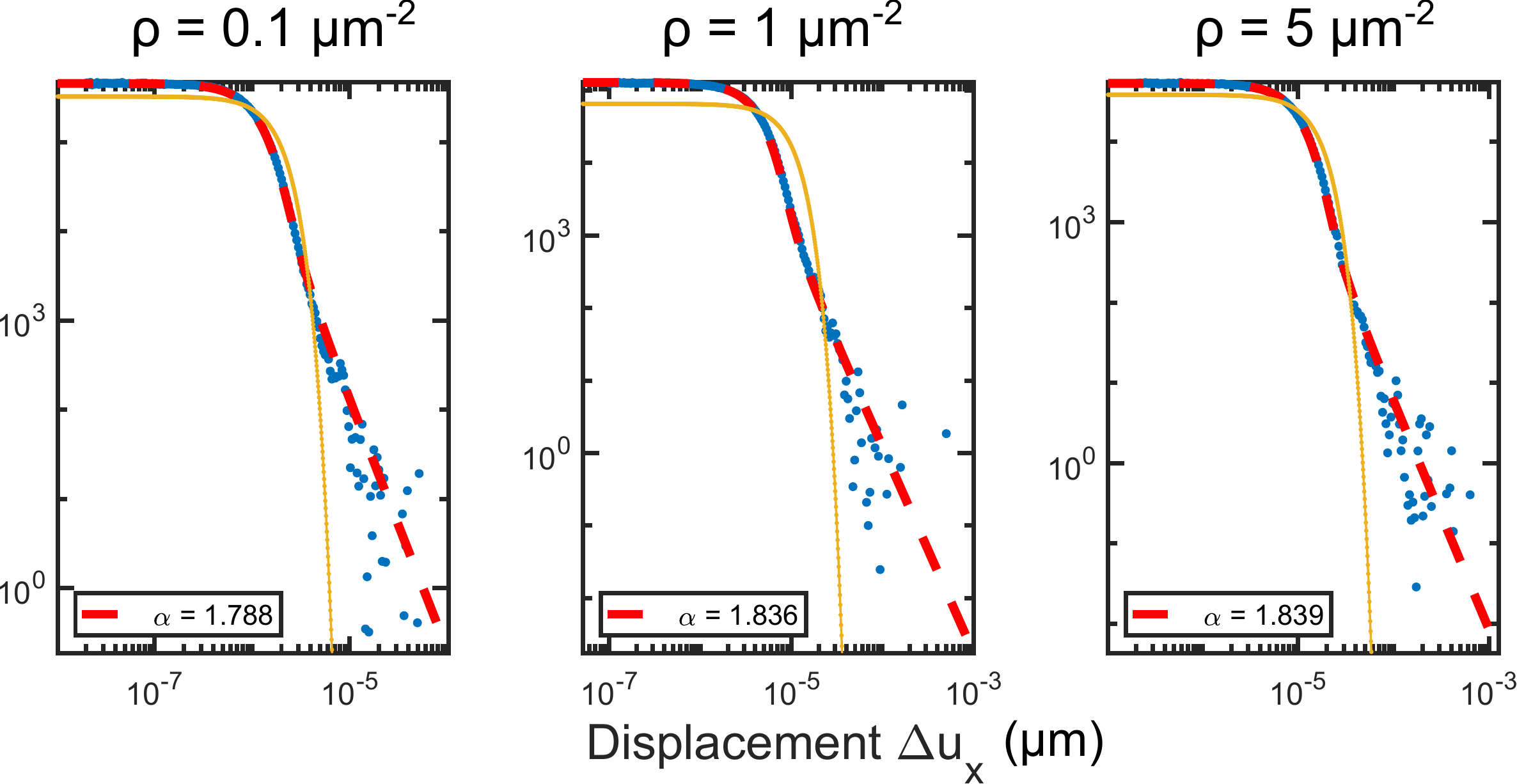}
    \caption{For small dipole sizes $d = 10^{-3} \um$, even when $G_{2D} = 1000 \textrm{Pa} \um$, we do not reach $\alpha = 2$. Van Hove plots at lag time $2 \tau \; (10 \textrm{s})$.}
    \label{fig:G2D_large_d1em3}
\end{figure}

Though it is not as apparent, the results in \fig \ref{fig:saffvanhove} also depend on the finite size of the dipole. When the dipole size is smaller, we tend to see $\alpha < 2$ even when $\lsd$ becomes large (\fig \ref{fig:G2D_large_d1em3}). 

\begin{figure}[h!]
    \centering
    \includegraphics[width=0.5\linewidth]{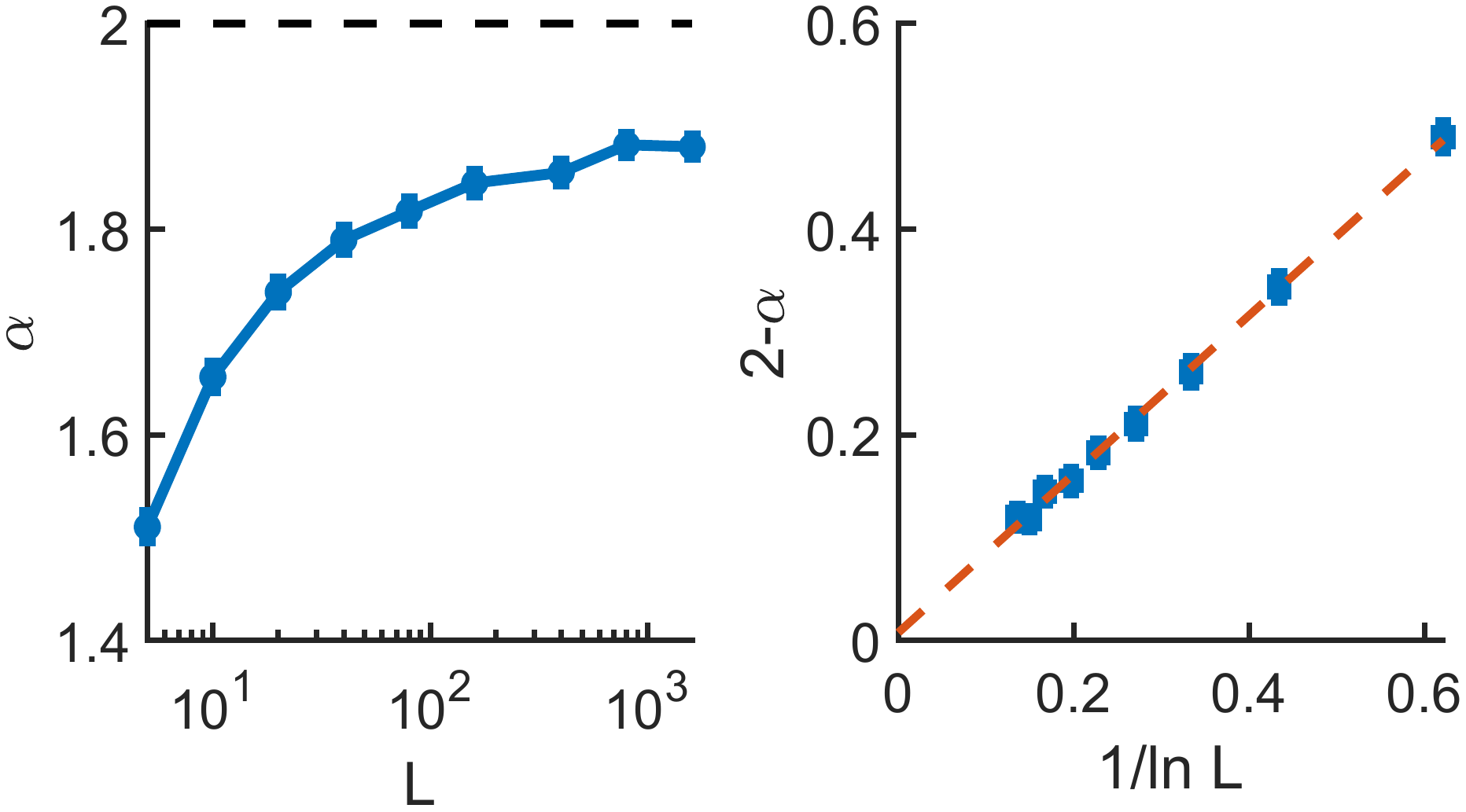}
    \caption{Our simple toy model with $D = 2$ and $m = 1$ has a systematic effect on $\alpha$ arising from the system size shows finite size effects compatible with $\alpha = \alpha_\infty - a / \ln L$.}
    \label{fig:toy_system_size}
\end{figure}

In principle, $\lsd$ becoming large should make the Oseen tensor logarithmic, and the response to an ideal dipole scale as $1/r$, leading to $\alpha = d/m = 2$. The deviation in \fig \ref{fig:G2D_large_d1em3} likely comes from a finite size effect, and is not arising from a special feature of the Saffman-Delbr\"uck tensor, but arises even in a toy model with $D = 2$ dimensions and a dipole response function with $m = 1$. We simulate a collection of ``toy dipoles" that generate a field $u(\rb) = c / r$, with $c$ a random number uniformly distributed between $-1/2$ and $+1/2$; we then generate $N$ random points over the $L \times L$ plane ($N$ chosen from a Poisson distribution with mean $\rho L^2$, $\rho = 0.1 \um^{-2}$), and compute the sum of the independent $u(\rb_i)$. This is then repeated $N_{\textrm{its}} = 5\times 10^4$ times. The resulting distribution can be fit to a \levy stable. We see the results in \fig \ref{fig:toy_system_size}. Though it might not be apparent from small changes in system size, $\alpha$ systematically increases toward $2$ as we increase the system size. We have been able to fit this form to $\alpha = \alpha_\infty - a / \ln L$ (\fig \ref{fig:toy_system_size}, right), with $a$ a constant. Extrapolating to $L = \infty$ suggests that we should reach $\alpha = 2$. In practice, we expect that in a physical system, the finite size may be quite relevant if the cell's Saffman-Delbr\"uck length is long, and ideal dipoles might see $\alpha < 2$. Interestingly, strong systematic effects of finite size simulation boxes are also seen in diffusion coefficients of membrane proteins and lipids in molecular dynamics simulations \cite{camley2015strong,vogele2016divergent}, where these effects similarly arise from the long-range nature of the Oseen tensor. 

\section{Effects of hard-core exclusion}
\label{app:cutoff}

\begin{figure}
    \centering
    \includegraphics[width = \figwidth]{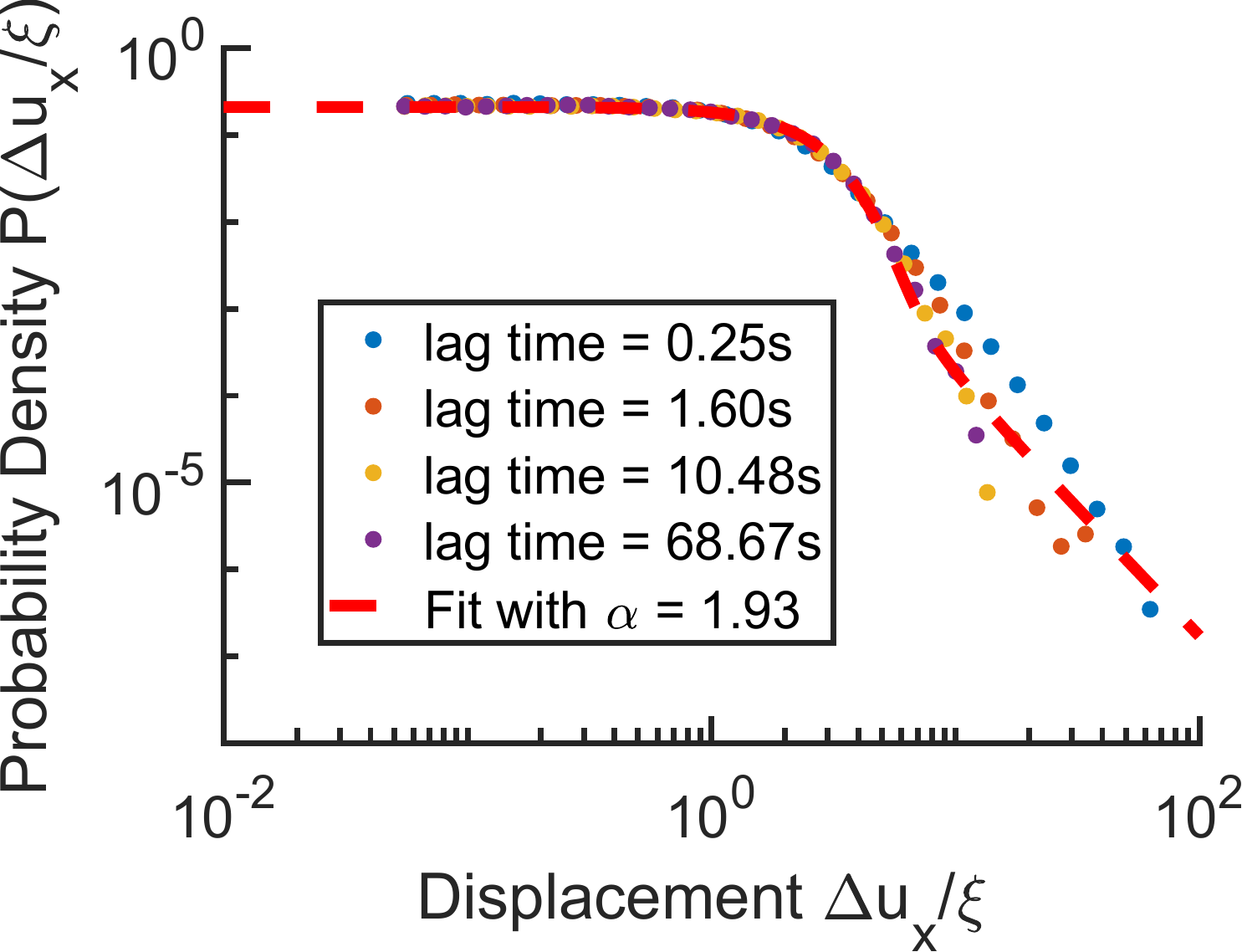}
    \caption{Effect of cutting off force dipoles from a distance $d$ from the origin. Van Hove distribution of displacement $u_x$ at different lag times, rescaled by a lag-time dependent factor $\xi(T) = \exp(\langle\log|\Delta u_x|\rangle)$. Other than the cutoff, parameters correspond to \fig \ref{fig:rescaledvanhove}.}
    \label{fig:cutoff_rescaledvanhove}
\end{figure}

\begin{figure*}
    \centering
    \centerline{\includegraphics[width=\textwidth]{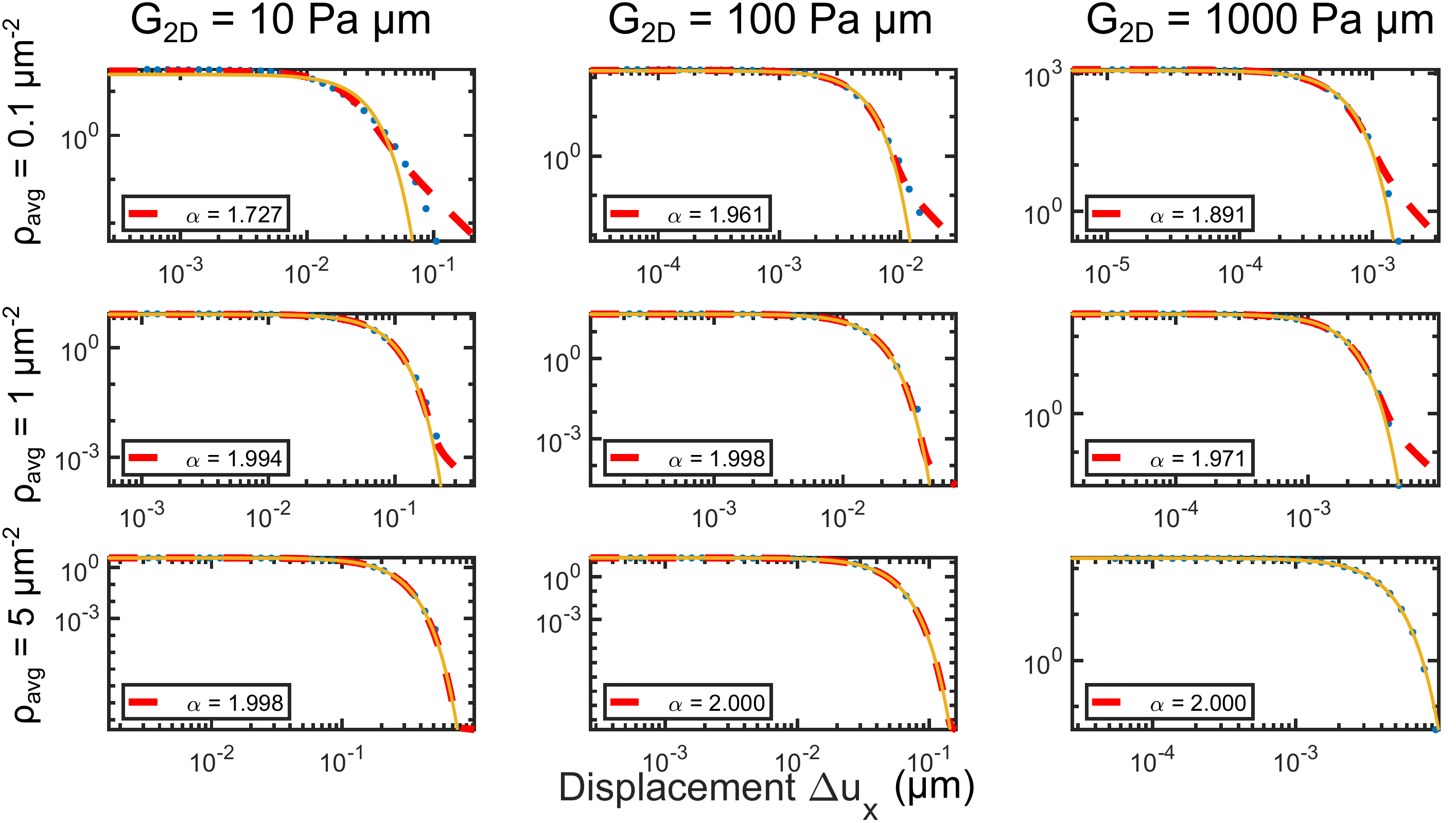}}
    \caption{Effect of cutting off force dipoles from a distance $d$ from the origin. Van Hove correlation plots $P(\Delta u_x)$ at lag time $2 \tau \; (10 \textrm{s})$ for different values of the cortex stiffness (and hence Saffman-Delbr\"uck length) and force dipole density. Blue dots are simulation, red dashed lines are best fit to \levy stable form, yellow solid lines best fit to a Gaussian ($\alpha = 2$). Other than the cutoff, parameters correspond to \fig \ref{fig:saffvanhove}.}
    \label{fig:cutoff_saffvanhove}
\end{figure*}

In the main paper, we assumed force dipoles are uniformly distributed over the entire system, including coming arbitrarily close to our tracer at the origin. This is an oversimplification that neglects the physical size of the force dipole. We show an extreme case of this exclusion in this appendix, not permitting force dipoles to come within a distance $d$ of the origin. (We reject these force dipoles and relocate them to a different location). These effects quantitatively, but not qualitatively change our conclusions. We present Van Hove curves corresponding to \fig \ref{fig:rescaledvanhove} and \ref{fig:saffvanhove} in the main text below as \fig \ref{fig:cutoff_rescaledvanhove} and \ref{fig:cutoff_saffvanhove}. 

%


\end{document}